\documentclass[twocolumn]{aastex63}

\newcommand{\ms}{{m s$^{-1}$}}
\newcommand{\cms}{{cm s$^{-1}$}}

\newcommand{\fap}{Fabry-P\'{e}rot}
\newcommand{\fsr}{$\Delta f_{\mathrm{FSR}}$}
\usepackage[utf8]{inputenc} 
\usepackage[T1]{fontenc}

\shorttitle{The HPF Fabry-P\'{e}rot Etalon}
\shortauthors{Terrien et al.}

\graphicspath{{./}{figures/}}

\begin{document}

\title{Broadband stability of the Habitable Zone Planet Finder Fabry-P\'{e}rot etalon calibration system: evidence for chromatic variation}

\correspondingauthor{Ryan Terrien}
\email{rterrien@carleton.edu}

\author[0000-0002-4788-8858]{Ryan C Terrien}
\affiliation{Carleton College, One North College St., Northfield, MN 55057, USA}

\author[0000-0001-8720-5612]{Joe P.\ Ninan}
\affil{Department of Astronomy \& Astrophysics,  525 Davey Laboratory, The Pennsylvania State University,  University Park, PA, 16802, USA}
\affil{Center for Exoplanets and Habitable Worlds,  525 Davey Laboratory, The Pennsylvania State University,University Park, PA, 16802, USA}

\author[0000-0002-2144-0764]{Scott A Diddams}
\affiliation{National Institute of Standards and Technology, 325 Broadway, Boulder, CO 80305, USA}
\affiliation{Department of Physics, University of Colorado, 2000 Colorado Ave, Boulder, CO 80309, USA}

\author[0000-0001-9596-7983]{Suvrath Mahadevan}
\affil{Department of Astronomy \& Astrophysics,  525 Davey Laboratory, The Pennsylvania State University,  University Park, PA, 16802, USA}
\affil{Center for Exoplanets and Habitable Worlds,  525 Davey Laboratory, The Pennsylvania State University,University Park, PA, 16802, USA}

\author[0000-0003-1312-9391]{Samuel Halverson}
\affil{Jet Propulsion Laboratory, California Institute of Technology, 4800 Oak Grove Drive, Pasadena, CA 91109, USA}

\author[0000-0003-4384-7220]{Chad Bender}
\affiliation{Steward Observatory, University of Arizona, 933 N Cherry Ave., Tucson, AZ 85721, USA}

\author[0000-0002-0560-1433 ]{Connor Fredrick}
\affiliation{National Institute of Standards and Technology, 325 Broadway, Boulder, CO 80305, USA}  
\affiliation{Department of Physics, University of Colorado, 2000 Colorado Ave, Boulder, CO 80309, USA}  

\author[0000-0002-1664-3102]{Fred Hearty}
\affil{Department of Astronomy \& Astrophysics,  525 Davey Laboratory, The Pennsylvania State University,  University Park, PA, 16802, USA}
\affil{Center for Exoplanets and Habitable Worlds,  525 Davey Laboratory, The Pennsylvania State University,University Park, PA, 16802, USA}

\author[0000-0002-7032-2350]{Jeff Jennings}
\affiliation{National Institute of Standards and Technology, 325 Broadway, Boulder, CO 80305, USA}
\affiliation{Department of Physics, University of Colorado, 2000 Colorado Ave, Boulder, CO 80309, USA}

\author[0000-0001-5000-1018]{Andrew J. Metcalf}
\affiliation{Space Vehicles Directorate, Air Force Research Laboratory, 3550 Aberdeen Ave. SE, Kirtland AFB, NM 87117, USA}

\author[ 0000-0002-0048-2586 ]{ Andrew Monson}
\affil{Department of Astronomy \& Astrophysics,  525 Davey Laboratory, The Pennsylvania State University,  University Park, PA, 16802, USA}

\author[0000-0001-8127-5775]{Arpita Roy}
\affiliation{Space Telescope Science Institute, 3700 San Martin Drive, Baltimore, MD 21218, USA}
\affiliation{Department of Physics and Astronomy, Johns Hopkins University, 3400 N Charles St, Baltimore, MD 21218, USA}

\author[0000-0002-4046-987X]{Christian Schwab}
\affiliation{Department of Physics and Astronomy, Macquarie University, Balaclava Road, North Ryde, NSW 2109, Australia }

\author[0000-0001-7409-5688]{Guðmundur Stefánsson}
\altaffiliation{Henry Norris Russell Fellow}
\affiliation{Princeton University, Department of Astrophysical Sciences, 4 Ivy Lane, Princeton, NJ 08540, USA}



\begin{abstract}
The comb-like spectrum of a white light-illuminated {\fap} etalon can serve as a cost-effective and stable reference for precise Doppler measurements. Understanding the stability of these devices across their broad (100's of nm) spectral bandwidths is essential to realize their full potential as Doppler calibrators. However, published descriptions remain limited to small bandwidths or short timespans. We present a $\sim6$~month broadband stability monitoring campaign of the {\fap} etalon system deployed with the near-infrared Habitable Zone Planet Finder spectrograph (HPF). We monitor the wavelengths of each of $\sim3500$ resonant modes measured in HPF spectra of this {\fap} etalon (free spectral range = 30 GHz, bandwidth = 820 - 1280 nanometers), leveraging the accuracy and precision of an electro-optic frequency comb reference. These results reveal chromatic structure in the {\fap} mode locations and in their evolution with time. We measure an average drift on the order of 2~cm~s~$^{-1}$~d$^{-1}$, with local departures up to $\pm5$~cm~s~$^{-1}$~d$^{-1}$. We discuss these behaviors in the context of the {\fap} etalon mirror dispersion and other optical properties of the system, and the implications for the use of similar systems for precise Doppler measurements. Our results show that this system supports the wavelength calibration of HPF at the $\lesssim10$~{\cms} level over a night and at the $\lesssim30$~{\cms} level over $\sim10$~d. Our results also highlight the need for long-term and spectrally-resolved study of similar systems that will be deployed to support Doppler measurement precision approaching $\sim10$~{\cms}.

\end{abstract}


\section{Introduction} \label{sec:intro}
High precision Doppler radial velocity (RV) measurements have progressed significantly in the past decade. Leveraging a suite of new technologies and analysis techniques \citep{2016PASP..128f6001F}, current state-of-the-art RV measurements have reached the $\sim$1 {\ms} precision regime on bright stars, and are pushing towards the $\sim$10 {\cms} precision (i.e.~one part in $3\times10^{9}$) required to detect an Earth-mass planet orbiting in the Habitable zone \citep{2013ApJ...765..131K} of a Sun-like star \citep{2020arXiv200706410D,2020AJ....159..187P}. New instruments and surveys striving for this level of precision include EXPRES \citep{2020AJ....159..238B}, ESPRESSO \citep{2010SPIE.7735E..0FP}, and NEID \citep{2018SPIE10702E..71S}. These and others are motivated by an array of timely scientific objectives and opportunities, including the drive for precise exoplanetary mass constraints to test hypotheses about exoplanet formation and evolution \citep[e.g.,][]{2019ApJ...878...36L} as well as to inform plans for future space-based studies of exoplanetary atmospheres \citep{2018PASP..130k4401K,2020PASP..132e4501F}.

Approaching the $\sim$10 {\cms} goal is a significant and multifaceted challenge \citep{2019BAAS...51g.232G}, which requires exquisite control of the spectrograph optomechanical platform \citep{2010SPIE.7735E..0FP, 2016ApJ...833..175S, 2019JATIS...5a5003R} and instrumental illumination \citep{2014SPIE.9147E..6BR, 2015ApJ...806...61H, 2018SPIE10702E..6FS, 2018SPIE10702E..6QK, 2020AJ....159..238B}, as well as robust analysis techniques \citep{2010SPIE.7735E..0FP,2020AJ....159..187P}. Even with state-of-the-art RV systems,  instrumental drifts\footnote{The term ``drift" is used here to describe the uncontrolled variations in a spectrograph or calibration system, which can manifest as changes in the measured spectra or derived quantities.} at the level of 10's of {\ms} or larger are routinely observed \citep[e.g.,][]{2020arXiv200601684B}. Careful wavelength calibration provides a path toward reducing the systematic noise floor imposed by these spectrograph drifts, by tracking a well-understood calibration source simultaneously with stellar observations, and using these to correct the instrumental drifts and isolate the stellar signal. 

Historically, wavelength references such as molecular iodine cells and Thorium-Argon (Th-Ar) lamps have provided stable and well-measured reference spectra for RV measurements, and have enabled RV measurements with a precision of $\sim$1 {\ms} \citep[e.g.,][]{2017AJ....153..208B,2019A&A...622A..37U}. However, the limited spectral coverage and irregular spectral line separations and amplitudes, as well as operational issues related to lifetime, chemical purity \citep[e.g.,][]{2018SPIE10704E..07N}, and commercial availability, have driven interest in new calibration sources. Among such systems are those based on laser frequency combs \citep[LFCs,][]{2007MNRAS.380..839M} and white-light illuminated {\fap} etalons \citep[FPs,][]{2010SPIE.7735E..4XW}, both of which yield stable optical spectra of regularly-spaced, approximately equal intensity, narrow spectral lines (resonant modes) that can be used for calibration. There are, however, substantial differences in the nature and utility of these two types of calibration sources.

The mode spectrum of an LFC is tied to precisely-known atomic radio frequency standards; the result is a well-defined optical frequency for each spectral line, with each line frequency known to a fractional absolute\footnote{i.e.~relative to the SI second.} precision limited only by the frequency standard.  For the best laboratory standards, a precision of $1\times10^{-18}$ has been achieved \citep{Diddamseaay3676}. However, for practical implementations with a GPS reference at an astronomical observatory, the typical precision is $1\times10^{-11}$ or better \citep{2019Optic...6..233M}. This level of absolute precision corresponds to an RV error of $<1$~cm~s$^{-1}$, making these systems promising for detection of Earth-mass planets over extended temporal periods. 

Although recent years have seen the deployment of a number of commercial and custom astronomical LFC systems \citep{2014SPIE.9147E..1CP,2019Optic...6..233M,2020arXiv200808949D,2020AJ....159..187P,2020arXiv200711013H}, these systems remain costly and operationally complicated. This is due largely to the confluence of two stringent requirements for use as Doppler calibrators: the need for broad (many 100's of nm) spectral coverage and the need to resolve individual LFC modes. A principal challenge is that maintaining flat illumination over such broad bandwidths---particularly towards blue ($<500$~nm) wavelengths---is difficult when coupled with the high repetition rate ($\gtrsim10$~GHz) required to generate resolvable comb modes at typical Doppler spectrograph resolution ($R = \lambda / \Delta \lambda $ typically between 50,000 and 100,000). For detailed discussion on these and other challenges, see \citet{ediss18701} and \citet{2017OExpr..2515058M}.

In comparison, FPs are simpler and less costly, and so offer an attractive alternative to LFCs. The core mechanism of an FP is an optical resonator formed by two partially-reflective mirrors, which is characterized by a spectrum of resonant longitudinal modes that are approximately equally-spaced in frequency \citep{1899ApJ.....9...87P}. This resonator can act as a stable filter of injected continuum, transmitting a regular, comb-like spectrum. Unlike in an LFC system, the center frequency of a given line in the spectrum of an FP is not known a priori, as the frequency of each line depends on the precise optical details of the FP, including its size, geometry, the characteristics of the mirror reflective coating, and how light is coupled in and out. Moreover, because the mode spectrum of an FP is tied to physical properties, and not atomic frequency standards, the frequencies of its modes may be influenced by environmental conditions \citep[e.g.,~temperature and pressure,][]{2017OExpr..2515599J}, as well as slow material drift \citep{1977Metro..13....9B}. 

Nonetheless, astronomical FP systems may be actively or passively stabilized; the stability and high information content of their spectra have been demonstrated to support $\lesssim1$ {\ms} RV precision in several systems \citep{2014PASP..126..445H, 2017JATIS...3b5003S, 2017A&A...601A.102C, 2018SPIE10702E..6AD, 2019A&A...624A.122C}. Further, an extensive history of use of FP etalon systems in the field of optical frequency metrology indicates that suitably designed FP systems can support narrow-bandwidth measurements ($<<1$nm) at a level of stability and precision far exceeding the 10 c{\ms} level over many years \citep[e.g.,][]{2009ApPhB..95...43D}. However, astronomical FPs are distinguished by their comparatively broad bandwidths ($\gtrsim500$ nm) and wide resonances\footnote{Equivalently, low values of the FP finesse.}; their use is therefore subject to different systematics. The long-term and spectrally-resolved broadband performance of astronomical FP systems at the $\lesssim10$ {\cms} level remains mostly undetermined. 

The FP analyzed in this work is a broadband (820-1280~nm), passively-stabilized astronomical FP. Our earlier laboratory study of this FP \citep[][hereafter J20]{Jennings2020} revealed significant deviations in the drift behavior of three FP resonant modes (at 780, 1064, and 1319 nm) compared to expectations if these drifts are caused by changes in the FP mirror separation (described in more detail in Section \ref{sec:fpdesc}). Confirmation of this behavior in situ at the observatory, as well as more spectrally extensive measurement, is essential if this and other FP systems are to be used to calibrate the long-term and potentially complicated drifts of the RV spectrographs targeting Earth-mass exoplanets. As an example, one possible implementation of an astronomical FP calibrator relies on the independent tracking of a small number of FP modes in order to track the drift the entire FP mode spectrum \citep[e.g.,~using Rb reference gas cells;][]{2014A&A...569A..77R,2018SPIE10702E..72S}. This approach may be limited by the extent to which the behavior of the full population of FP modes can be paramaterized by the behavior of this small set of modes. 

To help inform the use of this and other astronomical FP calibrators, we present here measurements extending over 6 months of the frequency stability of a broadband near-infrared (NIR) FP system for the Habitable-zone Planet Finder (HPF) spectrograph \citep{2014SPIE.9147E..1GM}. These measurements leverage a dedicated LFC calibration system \citep{2019Optic...6..233M} to probe the behavior of each of $\sim3500$ FP modes in the NIR. In Section \ref{sec:background} we provide descriptions of the spectrograph, LFC, and FP systems we use, and discuss the basic equations describing each. In Section \ref{sec:methods} we describe our dataset and methodology, highlighting the use of the LFC for wavelength calibration and the precision achieved in tracking the FP resonant modes. In Section \ref{sec:results} we describe the results of our measurements, showing the wavelength-dependence of the FP mode spectrum, the precision possible with the ``bulk" stability of the FP, and the significant chromatic structure in the drift rates of the FP resonant modes. In Section \ref{sec:discussion} we discuss the potential causes and implications of our results for Doppler spectrograph calibration. We summarize our key findings in Section \ref{sec:conclusion}.

\section{Background} \label{sec:background}
\subsection{The Habitable-zone Planet Finder} \label{sec:hpf}
HPF is a stabilized, fiber-fed, cross-dispersed echelle spectrograph at the Hobby-Eberly Telescope (HET) at McDonald Observatory in Texas, optimized for Doppler RV exoplanet detection around low-mass (M dwarf) stars. It observes in the near-infrared (820-1280~nm) over 28 echelle orders at high resolution ($R\approx 55,000$). HPF has demonstrated a thermal stability of $\sim1$~mK \citep{2016ApJ...833..175S} over $\sim1$~month timescales, and with a custom-built LFC calibration system (described in Section \ref{sec:LFC}) has been shown to be capable of a Doppler precision of $\sim1.5$~m~s$^{-1}$ on an M dwarf star over many months \citep{2019Optic...6..233M}. 

HPF is fed by three fibers: one corresponding to the science target, one corresponding to blank sky (to correct for emission lines in Earth's atmosphere), and one for the calibration source. For optimal drift correction during stellar observations, the calibration source may be observed simultaneously with the stellar target. Using the LFC to illuminate both the science and the calibration fibers has shown that these fibers track each other to $\lesssim 10$~cm~s$^{-1}$ \citep{2019Optic...6..233M}. 

Since its installation in 2017, HPF has been in regular use as a facility instrument at the HET, where its observations have enabled studies of nearby M dwarf planet companions \citep{2019Optic...6..233M,2020arXiv200611180S,2020AJ....159..100S,2020arXiv200614546K,2020arXiv200707098C,2020arXiv200712766S}, planetary atmospheres \citep{2020ApJ...894...97N}, stellar activity \citep{2020ApJ...897..125R}, and precise Doppler spectrograph systematics \citep{2019JATIS...5d1511N}.

\subsection{The HPF LFC system}
\label{sec:LFC}
A custom-designed electro-optic LFC calibration system is deployed with HPF. This system is described in detail in \citet{2019Optic...6..233M} and \citet{2019OptL...44.2673M}, but we briefly outline its properties here. A seed laser at 1064~nm is intensity and phase-modulated at 30~GHz and broadened in a silicon nitride waveguide to generate a spectral comb with 30~GHz mode spacing spanning the HPF bandpass. The output spectrum is flattened with a combination of passive and active filters. This system provides reliable calibration light for HPF on a nightly basis, and can be used simultaneously with stellar observations. The routine usage of the HPF LFC system is enabled by its exceptional reliability: the LFC was operational $>97\%$ of the time from May 2017 - Nov 2019. The accuracy and precision provided by LFC spectra are our primary reference for measuring the daily and long-term drift of HPF, and thus for determining the HPF wavelength solution at any given time (discussed in Section \ref{sec:wavecal}) that underlies the FP results presented here.

\subsection{The HPF {\fap} etalon system}
\begin{figure*}
    \epsscale{0.6}
\gridline{\fig{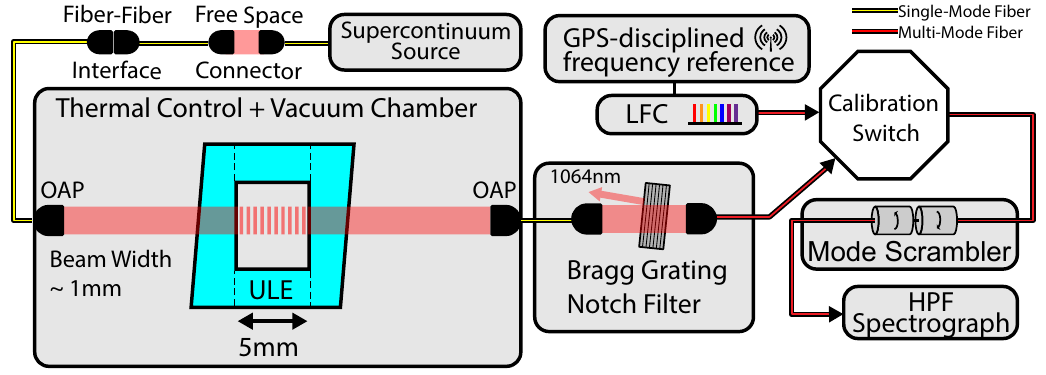}{0.6\textwidth}{(a)}
          \fig{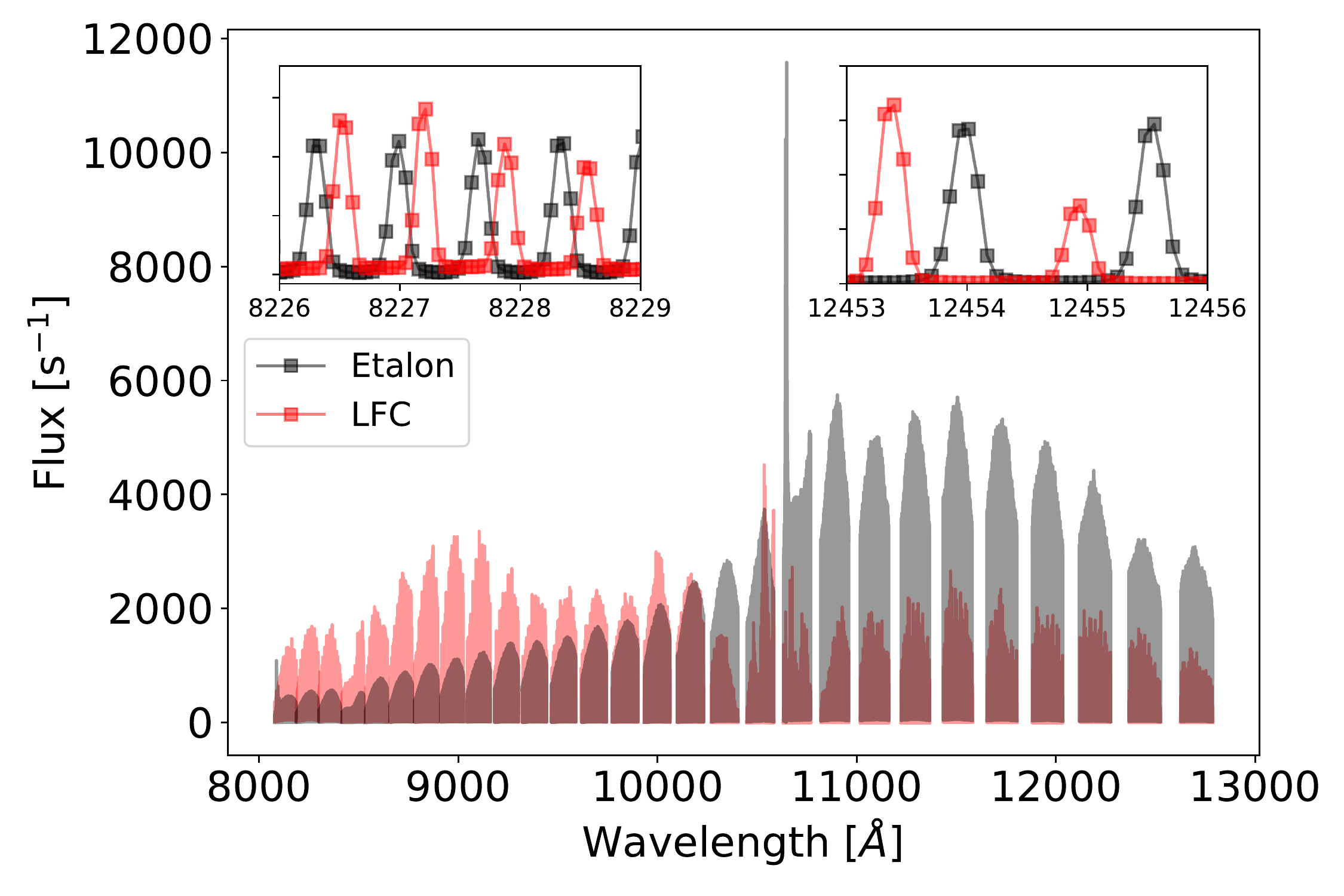}{0.4\textwidth}{(b)}
}
    \caption{(a) A schematic of the HPF FP calibration system. An NKT supercontinuum light source is passed through a short free-spacing connection and a fiber-fiber interface before illuminating the resonator, which is enclosed in a temperature-controlled vacuum chamber. Light is coupled into and out of fiber via off-axis parabolic collimators (OAPs). The resonator spacer and mirrors are made from ULE glass. The mirrors are wedged to suppress undesired parasitic etalons, and have an anti-reflection coating to suppress undesired reflection. The output light is filtered by a narrow-band Bragg grating filter, which suppresses the strong pump laser line from the NKT SuperK source to prevent saturation of the HPF detector. The FP light is one of several options that may be selected by the calibration switch to feed the calibration fiber of HPF. Alternately, the mode spectrum of the HPF LFC, traceable to the SI second via GPS, can be selected. The calibration light is passed through a mode scrambler prior to injection into the HPF spectrograph. (b) Example raw spectra of the HPF FP and LFC. The spectral variations result from a combination of the intrinsic source profile and the HPF throughput, and the shape within each order is dominated by the spectrograph blaze function. As shown with the rescaled spectra in the insets, both the LFC and FP modes are narrow and well-separated; there are $\gtrsim100$ modes in each of the 28 HPF echelle orders.}
    \label{fig:schematic}
\end{figure*}

The HPF FP is described in detail in \citet{Jennings2020}. Briefly, the FP resonator is constructed from two planar mirrors separated by a $L\approx5$~mm Ultra-Low Expansion (ULE) glass spacer (built by Light Machinery). This etalon is enclosed in a temperature-controlled vacuum chamber, and light is coupled in and out with single-mode optical fibers collimated by off-axis parabolic mirrors (housing and stabilization system built by Stable Laser Systems). The system is illuminated with an NKT SuperK supercontiuum source. The transmitted comb-like spectrum is once again collimated  and then filtered by a narrow-band Bragg grating notch filter (built by Optigrate, with thickness $\approx 2$~mm, FWHM~$\approx0.5$~nm, and $10^{-3}$ attenuation over $\sim1$~nm spectral width) in order to suppress the bright NKT SuperK pump laser wavelength near 1064~nm. The filtered light is coupled into a multimode fiber and injected into the HPF calibration switch. From there it passes through a mode scrambler (built by Giga Concepts) and finally into the HPF spectrograph. A schematic sketch of the system is shown in Figure \ref{fig:schematic}.

The fundamental parameters of the optical spectrum of this FP system are its free spectral range and finesse. The free spectral range is approximately 30~GHz, corresponding to approximately six resolution elements of HPF, and the free spectral range is explored in much more detail below. The targeted (coating-limited) finesse across the HPF bandpass is $\approx40$; measured (in the lab at NIST) finesse values are 33 (780~nm), 43 (1064~nm), and 41 (1319~nm) \citep{Jennings2020}. These correspond to an intrinsic line width of approximately 800~MHz, a factor of $\sim7$ narrower than the instrumental profile of HPF. Due to the uncertainties involved in de-convolving these effects, an independent measurement of the finesse (or the stability thereof) is beyond the scope of this work.

\subsection{Theoretical {\fap} Description}
\label{sec:fpdesc}
We discuss here the salient theoretical aspects of a broadband planar FP with mirrors separated by a distance $L$. Collimated light is coupled in through the partially-transmissive mirrors, and successive reflections lead to interference and resonance for certain frequencies. Notably, the dielectric mirrors on the inside of the resonator in the HPF FP are designed for reflectivity of $\approx92$\% across the entire 820-1280~nm band (corresponding to finesse of $\approx40$). These mirrors achieve high reflectivity by intereference off the different layers in the dielectric stack; maintaining such a reflectivity over the full 820-1280~nm band requires many layers and so the full mirror coating has a non-negligible thickness. Solutions to Maxwell's equations for this situation are well-studied; modifications to the resonance condition \citep[as in][]{1988PhRvA..37.1802D} yields a frequency-dependent expression for the FP longitudinal mode spacing (free spectral range, \fsr):
\begin{equation}
    \label{eq:fsr}
    \Delta f_{\mathrm{FSR}}(f) = \frac{c}{2nL + \frac{c}{\pi} \frac{\partial \phi(f)}{\partial f}},
\end{equation}
where $f$ is the frequency of the light, $n$ is the refractive index of the propagation medium, $\phi(f)$ is the frequency-dependent phase delay from the mirror reflection, and $c$ is the speed of light.

An often-used simplification of Equation \ref{eq:fsr} is notable: if phase delays from the mirrors are neglected and the mirrors are in a medium with refractive index $n$ (presumably a slow function of frequency), {\fsr} is simply
\begin{equation}
    \label{eq:FSRsimple}
    \Delta f_{\mathrm{FSR}} = \frac{c}{2 n L}.
\end{equation}
For mode index $m$, the corresponding frequency would then be 
\begin{equation}
    \label{eq:freqFSR}
    f_m = m \; \Delta f_{\mathrm{FSR}}.
\end{equation}
The most fundamental parameter describing the FP mode spectrum is the mirror separation $L$; Equations \ref{eq:FSRsimple} and \ref{eq:freqFSR} suggest that changes in $L$ should drive mode frequency changes, and the ratio of such changes between two modes should be equivalent to the ratio of the mode indices (see discussion in J20).

\section{Methods} \label{sec:methods}
We measure and track the centroids of each of $\sim3500$ modes of the FP spectrum in HPF spectra obtained over $\sim1000$ epochs spanning six months. An accurate wavelength calibration (described in Section \ref{sec:wavecal}) allows us to translate these centroids to wavelengths, correcting for the spectrograph drift. The resulting data provides a spectral and temporal baseline to assess the wavelength-dependent stability of the FP spectrum.

\subsection{Dataset}
The HPF FP is part of a suite of routinely-used calibration sources, which also includes the LFC and hollow cathode lamps. Daily morning and evening calibration sequences each include between one and four observations of the FP, with a uniform integration time of $\approx160$~s, resulting in a total number of recorded counts that is consistent to $\lesssim 10\%$ across observations. These FP observations are typically taken immediately following observations of the LFC, resulting in a minimal spectrograph drift for cross-calibration between the LFC and FP.

The one-dimensional spectra used here are optimally extracted \citep{1986PASP...98..609H} from two-dimensional images from the H2RG detector array \citep{2012SPIE.8453E..10B} using a custom HPF data reduction pipeline, which provides flux and variance measurements for each pixel and is described in detail in the supplemental material of \citet{2019Optic...6..233M} and in \citet{2018SPIE10709E..2UN}. Each FP spectrum has an average S/N (signal-to-noise ratio per extracted pixel at mode peaks) of $\approx 490$, and an approximate total photon-limited RV precision \citep{2001A&A...374..733B} of 4.9~{\cms}.

We select for this study a period spanning November 2018 through May 2019, which avoids spectrograph and LFC maintenance events and spans a period of uniform FP drift behavior. Later epochs do show some deviations from the behavior presented here, which are under study. Including all the FP spectra taken in the November 2018 - May 2019 timeframe yields 1050 spectra.

\subsection{Wavelength Calibration of FP spectra}
\label{sec:wavecal}
The HPF LFC provides an absolute calibration of the entire HPF spectrum (i.e.~an accurate and precise mapping of pixel to wavelength or frequency in an extracted spectrum--the wavelength solution) at any given time. However, the LFC is not continuously observed, and \textit{cannot} be observed simultaneously with the FP, so the determination of the wavelength solution for an arbitrary epoch hinges on a reliable understanding of the drift of the HPF wavelength solution. Careful daily measurements of the LFC and FP over $\sim$ 2 years have enabled the construction and refinement of an accurate multi-parameter model for this drift, the dominant terms of which correspond to a uniform pixel offset, a magnification term, and offsets affecting each readout channel of the H2RG detector. Here we briefly outline the relevant aspects of the wavelength calibration for FP spectra; a complete description of the HPF wavelength calibration algorithm and performance will be published in a later paper.  

The HPF wavelength solution and drift measurement begins with a high S/N LFC template, which is created by combining multiple LFC spectra that are close in time. These LFC frames are combined using an estimate of the underlying drift which is iteratively refined. The frequencies of the LFC modes in this template are known from the comb equation:
\begin{equation}
    \label{eq:comb}
    f_m = m f_r + f_0,
\end{equation}
where $m$ is the LFC mode index, $f_r$ is the LFC repetation frequency, and $f_0$ is the LFC offset frequency \citep{Diddamseaay3676}. In the HPF LFC system, both $f_r$ and $f_0$ are radio frequencies that are tracked with respect to a GPS-disciplined reference oscillator. The mode index $m$ is easily determined by coarse comparison of the template to the spectrum of a hollow cathode lamp. Thus, each mode frequency in the template spectrum is absolutely determined. The corresponding pixel position is determined by a least-squares fit of a Gaussian function to each LFC mode in this template, which provides $\sim100$ anchor points for the wavelength solution across each of the 28 echelle orders of HPF. The wavelength solution for this template is then taken as the least-squares fit of a fifth-order polynomial, fit independently to these anchor points in each spectral order. The residuals to these fits are largely normally distributed with a scatter of $\sim10$~{\ms}, although for approximately a third of the orders there is some correlated noise at the level of 10-20~{\ms} at scales of tens of modes. 

This template wavelength solution can then be appropriately transformed by a drift model to account for the instrumental drift to any epoch of interest. This model has been iteratively refined over time to represent the drift of HPF with the fewest possible parameters to maximize the S/N of the drift measurement. Each new LFC observation is compared to the LFC template in the context of this model to determine the appropriate value for each drift model parameter. The drift model addresses only changes in the ``dispersion'' sense, i.e.~those that are accessible in the one-dimensional extracted spectra.

Our HPF drift model comprises a second-order polynomial function of pixel index within each order, along with discrete offsets for the four read-out channels of the H2RG detector, which were found to drift apart slowly. Over short (daily) timescales, the dominant effect is seen in the $0^{\mathrm{th}}$-order coefficient of the polynomial. Changes in this coefficient are equivalent to a linear offset of the spectrum projected onto the H2RG detector array. This coefficent shows a daily sawtooth pattern, with a peak-to-peak amplitude corresponding to approximately 15 {\ms} \citep[as shown in Figure A1 of][]{2020AJ....159..100S}. This pattern is thought to be driven by the change in mechanical loading from the daily fill and subsequent evaporation of the liquid nitrogen coolant used by HPF, which is stored in a tank suspended underneath the HPF optical bench. The highly repeatable shape of this sawtooth allows us to accurately interpolate the appropriate value for the $0^{\mathrm{th}}$-order coefficient at any epoch.

The $1^{\mathrm{st}}$ and $2^{\mathrm{nd}}$-order coefficients in our drift model polynomial vary smoothly and slowly, on the timescale of weeks. The offsets for each readout channel also vary slowly, on an even longer timescale. This behavior allows the accurate interpolation of the relevant parameters in our drift model for any given epoch. Combined with the dominant term above, this establishes a precise determination of the wavelength solution for any single spectrum.

\subsection{Mode Measurement}
\label{sec:modemeasurement}
\begin{figure*}
\gridline{\fig{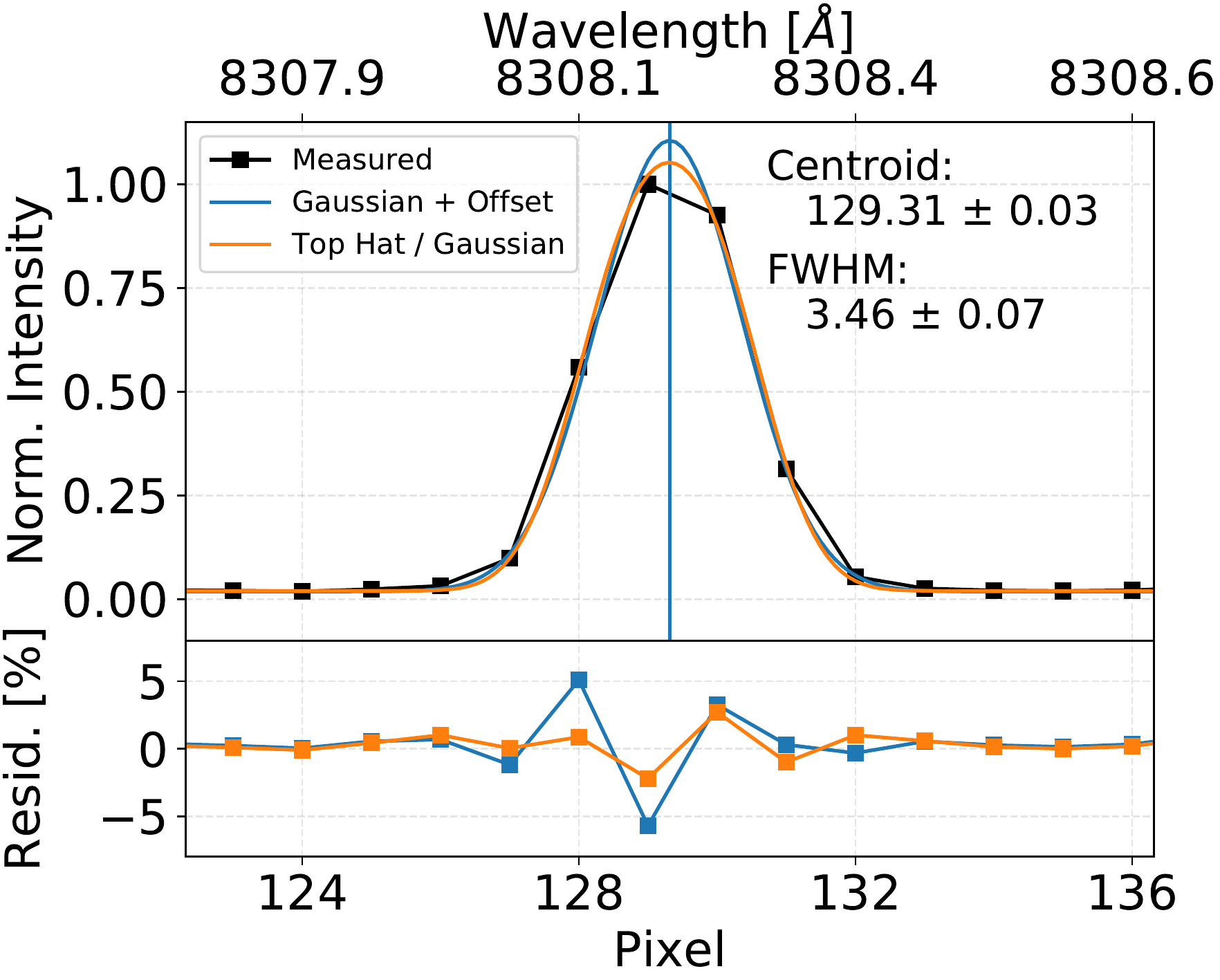}{0.3\textwidth}{(a)}
          \fig{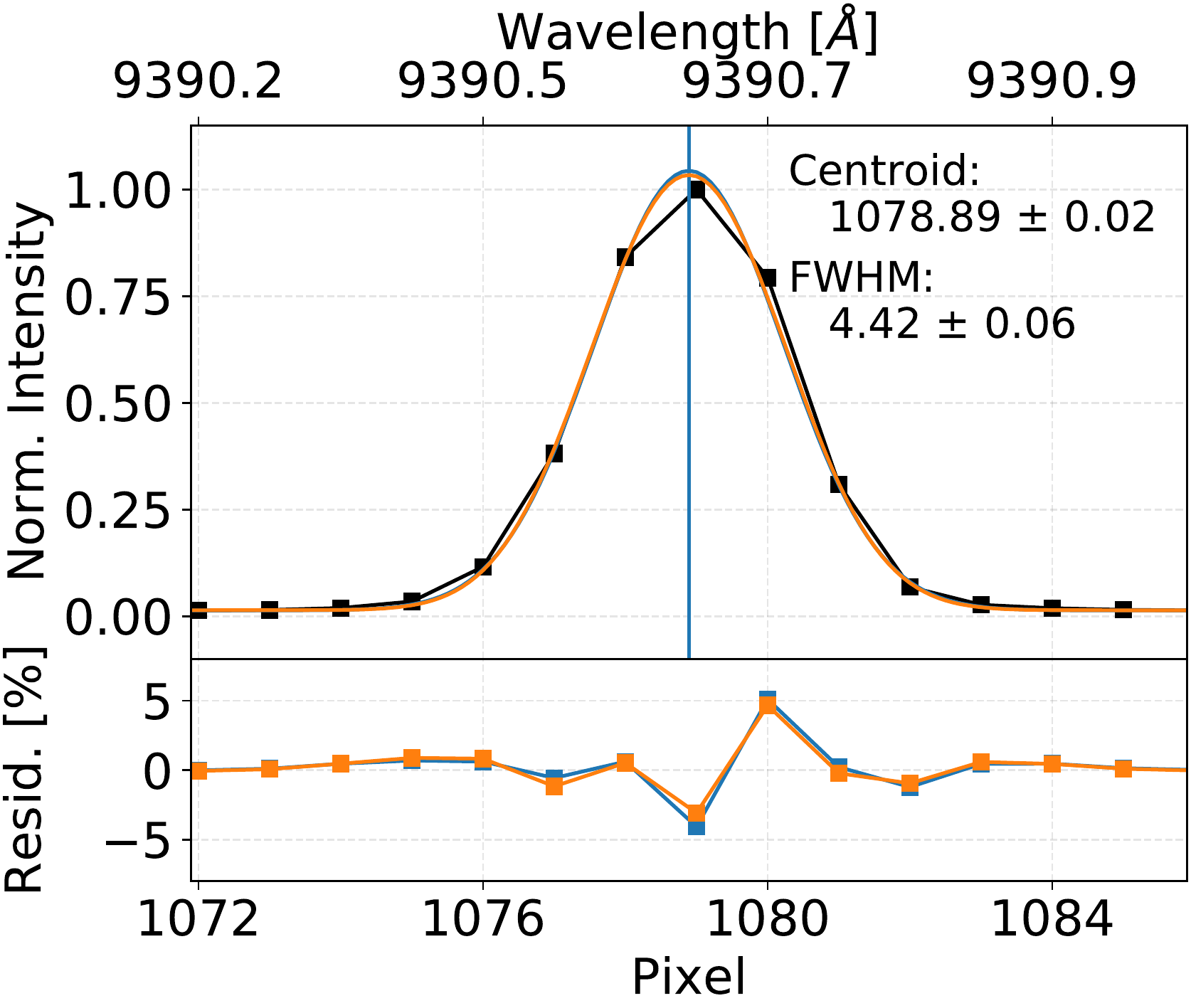}{0.3\textwidth}{(b)}
          \fig{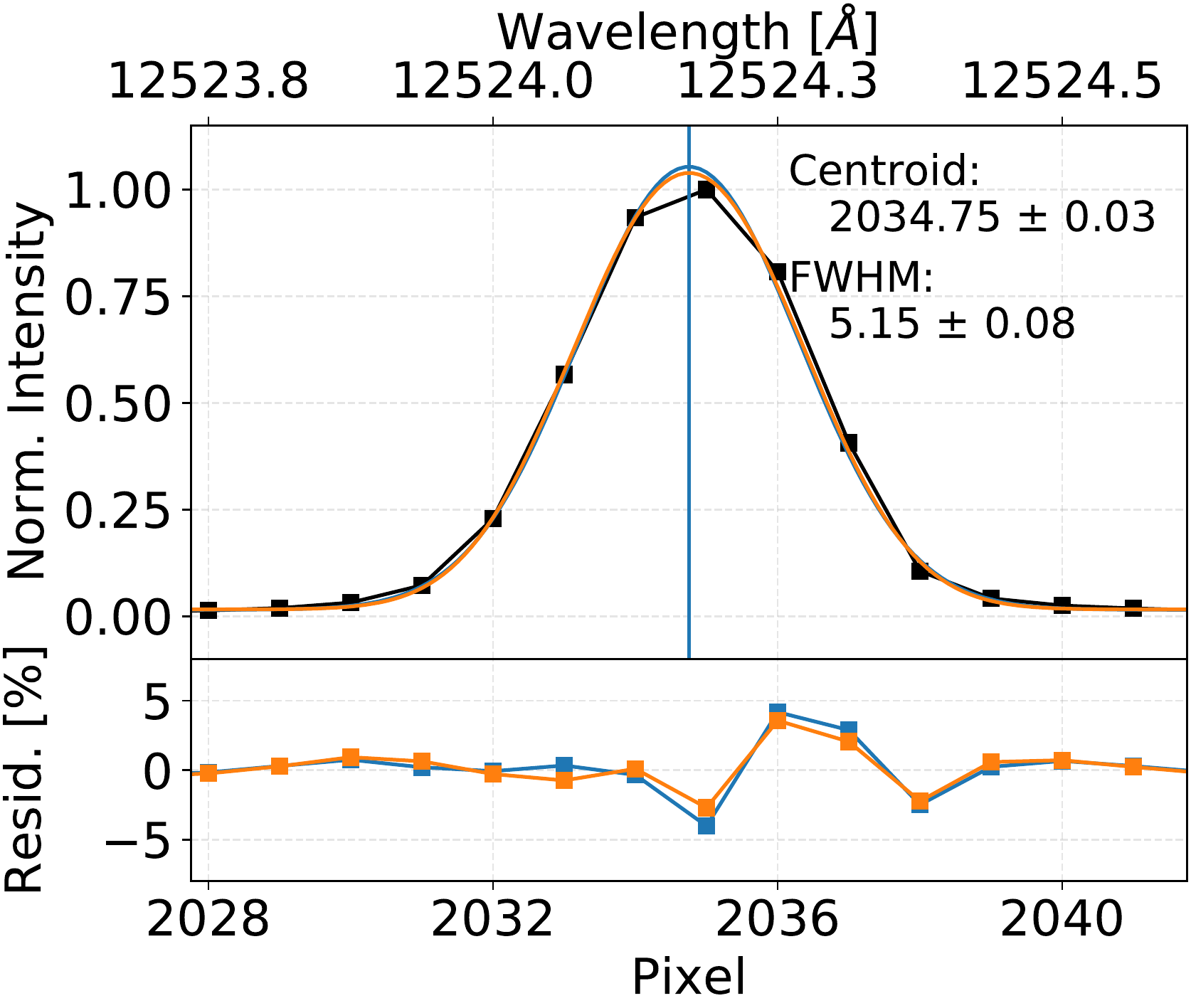}{0.3\textwidth}{(c)}
          }
\caption{Example fits for three modes across the HPF bandpass. We tested the performance of a simple Gaussian + offset model as well as the convolution of a top hat and Gaussian, and found only marginal reductions in the residuals with the more complicated model. We elect to use the simple Gaussian + offset, which has typical residuals $\lesssim 5\%$.
\label{fig:fits}}
\end{figure*}

Within each FP spectrum, we use the corresponding wavelength solution to measure the central wavelength of each individual FP resonant mode. We measure the centroid of each  mode independently using a least-squares fit of a Gaussian with a constant offset, a robust technique that provides easy-to-interpret outputs. A representative set of these fits is shown in Figure \ref{fig:fits}. The fit is performed on a 16-pixel window around each mode peak, which encompasses the full width of the $\sim5$ pixel line width and surrounding baseline. We note that these line widths are dominated by the instrumental profile, as the intrinsic FP line width is $\sim7$ times smaller. The residuals are typically $<5\%$.

As indicated by the structure in the residuals of Figure \ref{fig:fits}, the FP line profiles are not perfectly well-represented by a Gaussian. The fibers illuminating HPF are masked by a slit; the line profile is therefore expected to be slightly more flat-topped than a Gaussian. Further, the line profile contains some asymmetries resulting from optical aberrations impacting the HPF response function. Similar fits to the LFC spectra (whose modes are intrinsically even narrower than the FP modes) reveal equivalent structure in the residuals, indicating that this behavior is not unique to the FP.

To probe the sensitivity of our results to choice of fitting function, we repeated our analysis with a more complicated model corresponding to a convolution of a Gaussian and a top-hat function. While the residuals were reduced, the improvement was generally marginal and did not justify the increased complexity of the model. We therefore favor the simpler offset Gaussian model, and are careful to measure mode characteristics and behavior across groups of modes in order to smooth over variations that may relate to the aberration-related asymmetries (which correlate across echelle orders) and limited sampling of individual modes.

We further mitigate the potential impact of the unique systematics for any individual mode fit by focusing in this work on the patterns that are coherent across tens to hundreds of modes and across HPF spectral orders. All of our results below involve averaging over tens to hundreds of modes.

\subsection{Centroid Precision} \label{sec:precision}
An accurate estimate of the measurement precision on a single FP mode is necessary for determining the significance of spectral or temporal variations observed. To help understand the precision obtained with our data, we consider the fundamental photon-limited precision ($\delta v$) on a single mode, which can be estimated as in \cite{1987AcMik...3..215B}:
\begin{equation} \label{eq:dv}
    \delta v \approx A \frac{\mathrm{FWHM}}{\mathrm{S/N} \times \sqrt{n}},
\end{equation}
where $A$ is a (line shape-dependent) constant of order unity \citep{2007MNRAS.380..839M}, FWHM is the full-width half max of the mode, S/N is the peak signal-to-noise ratio\footnote{For FP modes measured with HPF, the ratio between integrated S/N and peak S/N is approximately 0.5.} in the mode, and $n$ is the number of pixels sampling the mode. 

\begin{figure*}
\gridline{\fig{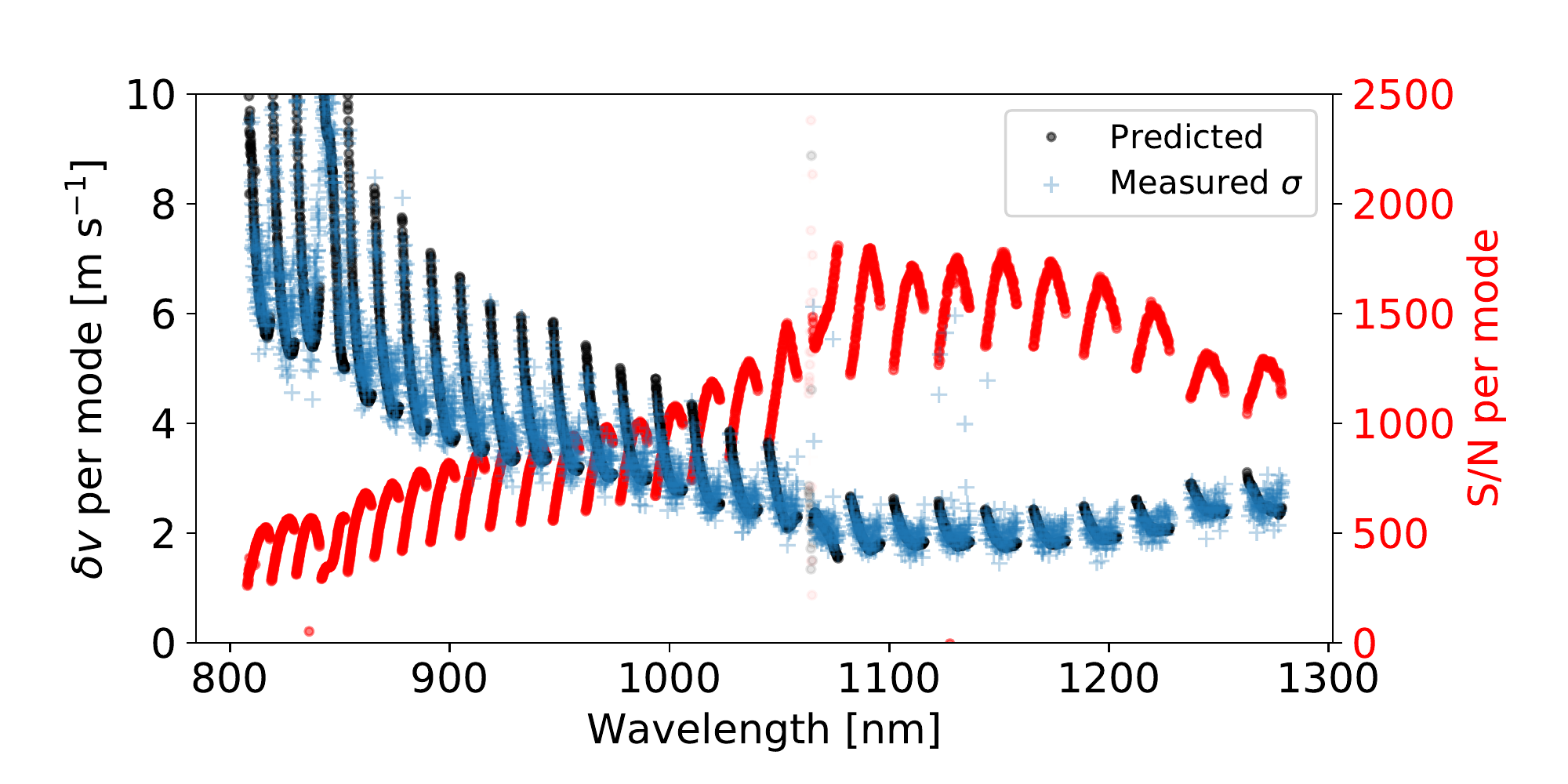}{0.6\textwidth}{(a)}
          \fig{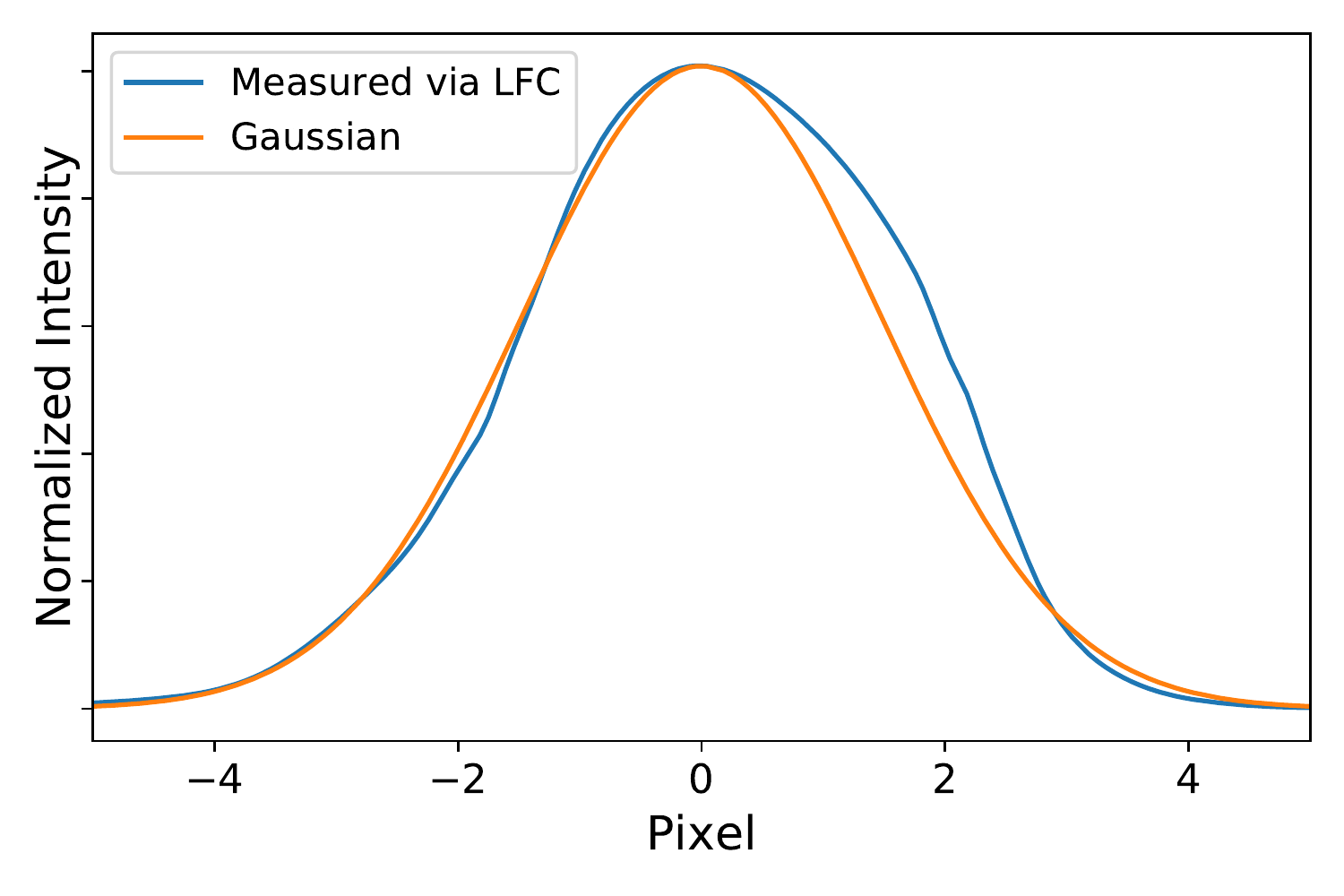}{0.4\textwidth}{(b)}
          }
    \caption{(a) The S/N and corresponding predicted mode centroid precision for all modes in a single 160~s observation of the HPF FP spectrum, as derived using Equation \ref{eq:dv}. The S/N values are photon noise-dominated, and spectral variations in the overall FP flux distribution are dominated by both the output spectrum of the SuperK source and the instrumental efficiency of HPF. Also shown are measured standard deviations ($\sigma$) for each mode centroid spanning 100 epochs from a typical period from April 24, 2019 through May 14, 2019. Points within {1~\AA} of the 1064~nm pump laser are deemphasized because they are subject to instability resulting from the intensity variations of the pump laser. (b) Two example profiles used to measure the scaling factor $A$ in Equation \ref{eq:dv}; a simple Gaussian (the functional form used for the fits described in Section \ref{sec:modemeasurement}) and one measured directly using the HPF LFC, corresponding to the red edge of echelle order index 5 (approximately 8765\AA). Both profiles yield an estimate of $A\approx 0.4$.}
    \label{fig:snrs}
\end{figure*}

We follow \citet{2007MNRAS.380..839M} to confirm the value for $A$ for FP modes in HPF spectra: we repeatedly simulate and recover the centroid of a mode with appropriate sampling and noise characterized by Poisson statistics and a small amount ($\sim5$~electrons) of detector read noise. Performing this numerical experiment with a Gaussian line at a variety of widths and S/N values, we confirm the estimate in \citet{2007MNRAS.380..839M} that $A \approx 0.4$. We repeated this experiment with a high-resolution line profile measured by scanning the HPF LFC \citep[similar to e.g.,~][]{2012MNRAS.422..761M} as in \citet{2019JATIS...5d1511N}. The tested profile corresponds to the edge of the HPF detector, where optical aberrations are maximized, and is shown in Figure \ref{fig:snrs}. The resulting estimate for $A$ was still approximately 0.4, confirming that this value is appropriate across the HPF spectrum of the FP.

We estimate S/N for a given mode using the flux and variance outputs of our HPF data reduction pipeline. A typical distribution of S/N values for a single 160~s exposure and the resulting photon-limited uncertainties (related through Equation \ref{eq:dv}) are shown in Figure \ref{fig:snrs}, along with the measured standard deviation ($\sigma$) for each mode centroid across 100 epochs during a typical period confirming that our measurements are mostly photon noise-limited for a single epoch. 

\section{Results} \label{sec:results}
\subsection{Static Description}
\label{sec:static}
The starting point for understanding the temporal behavior of the HPF FP mode spectrum is a precise description of its properties at a single point in time. Besides defining a reference point for measuring the temporal behavior of the FP mode positions, examining the measured {\fsr} may provide clues as to whether the HPF FP mode spectrum is measurably affected by phase-related deviations as suggested by Equation \ref{eq:fsr}.

\begin{figure*}
    \plotone{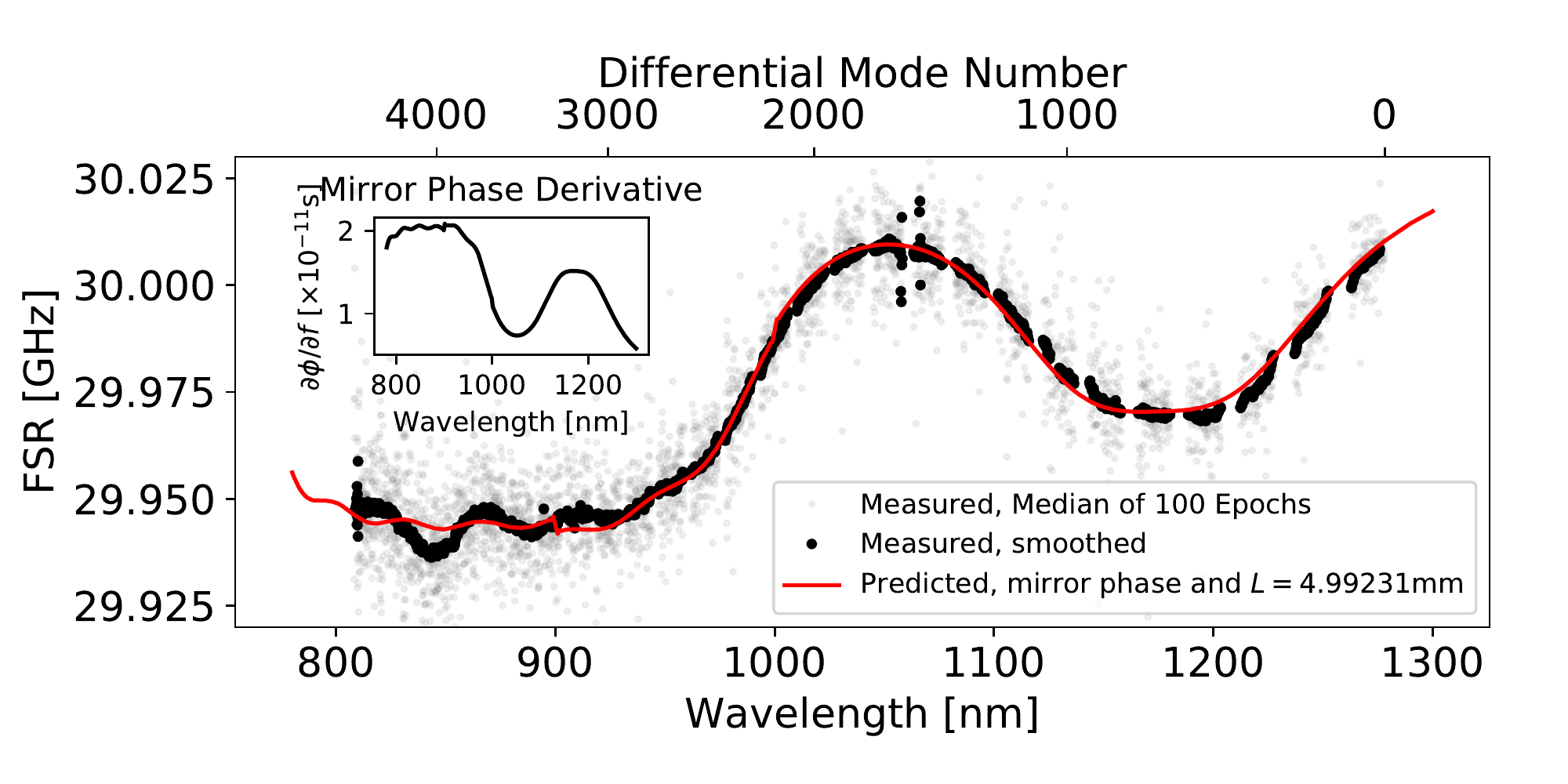}
    \caption{The FP free spectral range (\fsr) derived from the representative 100 epochs spanning April 24, 2019 - May 14, 2019. FP modes are identified by their wavelengths and ``Differential Mode Number" relative to the lowest mode index in our spectrum. Short-range variations are smoothed with a 31-point rolling (spectral) mean. The ``bulk'' value of approximately 30 GHz primarily results from the cavity spacing, while spectral variations can be understood as resulting from wavelength-dependent mirror phase delays. Projecting the (vendor-provided) phase delays for our mirrors through Equation \ref{eq:fsr} suggests that the {\fsr} variations are largely related to the mirror properties.}
    \label{fig:fsrs}
\end{figure*}

We establish an accurate ``single-epoch" description of the spectrum by averaging over the results of our mode centroid measurements from the representative period from April 24, 2019 through May 14, 2019 (the same period used in Figure \ref{fig:snrs}). The resulting {\fsr} measurements are shown in Figure \ref{fig:fsrs}. The measured {\fsr} values cluster near the specified value of 30 GHz, but variations of approximately 50 MHz are clearly detected over large (tens-hundreds of modes) ranges. 

We note that short range (mode-to-mode) structure is also detected, which can be seen as the scatter in the ``raw'' points plotted in Figure \ref{fig:fsrs}, which is elevated above the $\sim$MHz-level uncertainty expected for a pair of mode centroids (see Section \ref{sec:precision}). The origin of this short-range structure is difficult to trace, because any individual mode fit is subject to unique systematics due to detector effects (e.g., persistence and quantum efficiency variations) and the varying mismatch between the ``true'' line profile and the uniform Gaussian used (see Section \ref{sec:modemeasurement}. We focus instead in this work on the behaviors which are coherent across tens to hundreds of modes and across HPF spectral orders, which is more interpretable with our current understanding of HPF systematics.

The spectral shape of the FSR variations shown in Figure \ref{fig:fsrs} can be connected to the phase delay curve of the FP mirror using Equation \ref{eq:fsr}. Using this equation and the vendor-provided phase curve for the FP mirror (shown in the inset of Figure \ref{fig:fsrs}, we perform a least-squares fit to the measured {\fsr} values and find $L=4.99231 \pm 0.00003$~mm.\footnote{This fit assumes no uncertainty in the phase curve and should not be understood as an accurate determination of $L$. Subtleties resulting from mirror dispersion in the use of {\fap} etalons for precise length measurements are discussed in detail in \citet{1985JOSAA...2.1869L,1986JOSAA...3..909L}.} The good correspondence between the {\fsr} values predicted by this $L$ (which sets the bulk offset of the {\fsr}) and the phase curve (which imposes the spectral shape) indicates that most of the observed variation results from the mirror phase.

\subsection{Temporal Monitoring}

\begin{figure*}
\gridline{\fig{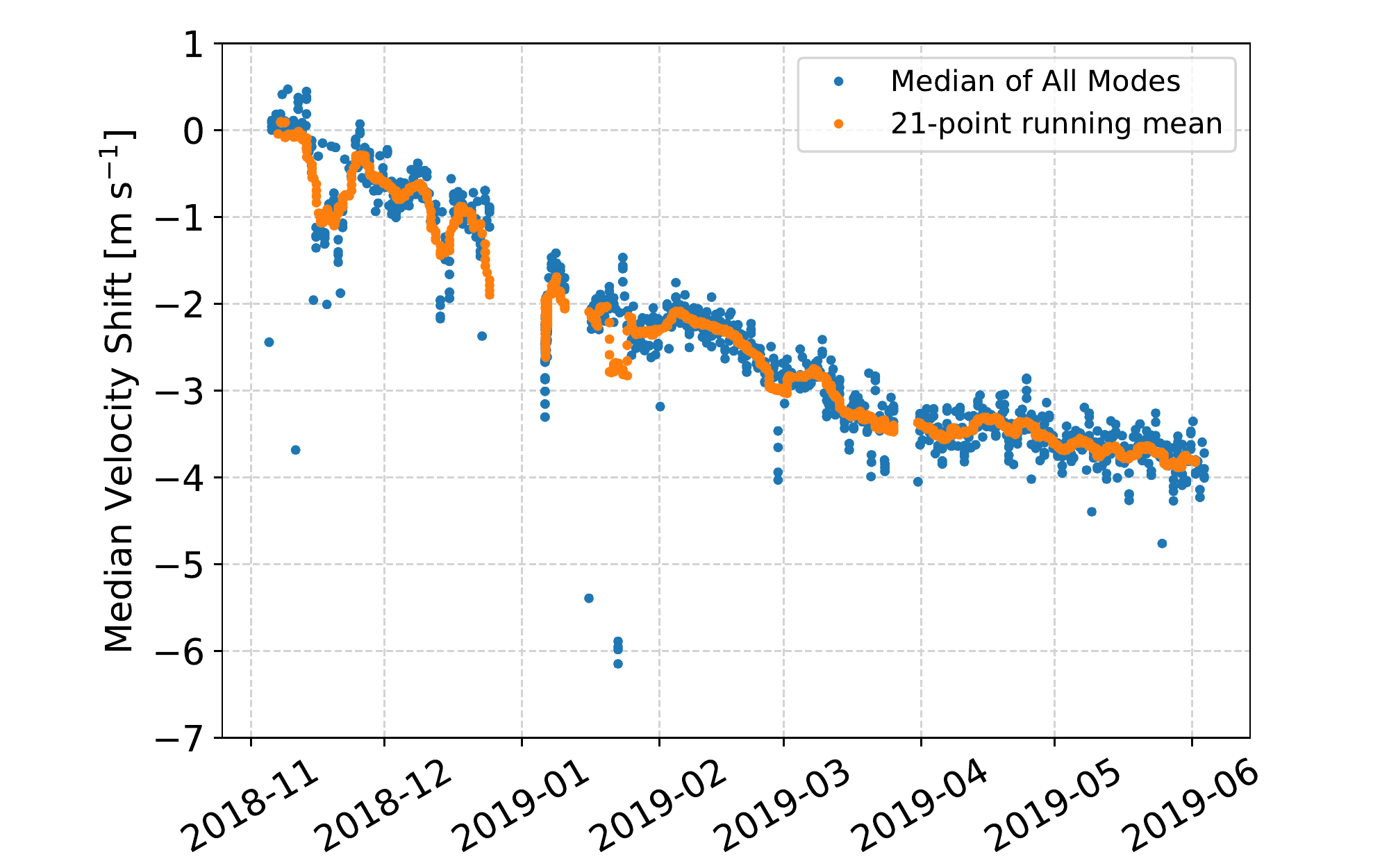}{0.5\textwidth}{(a)}
          \fig{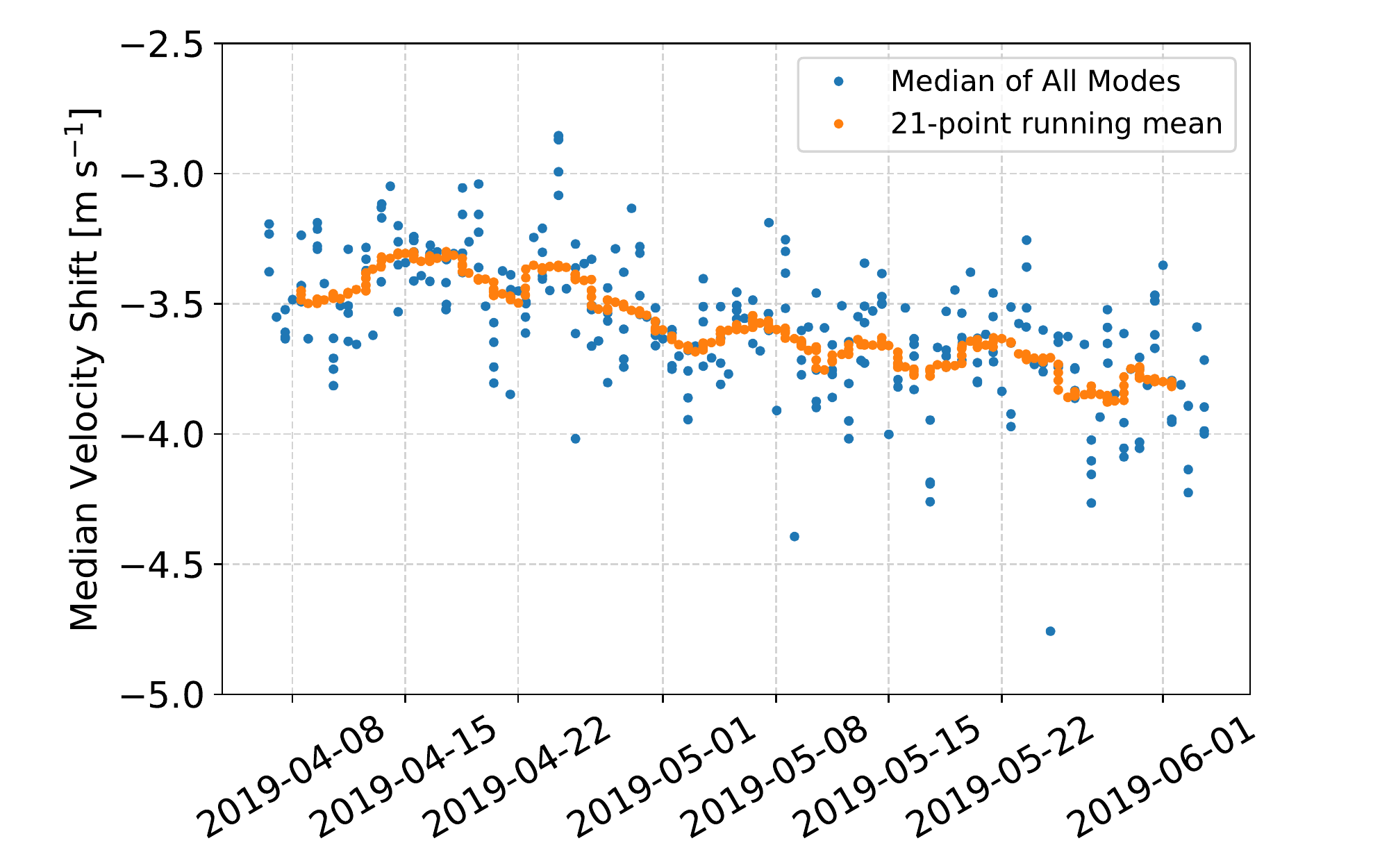}{0.5\textwidth}{(b)}
          }
\caption{The absolute drift of the FP, represented as a bulk velocity shift. Panel (a) shows the full timespan presented in this work, including points where the stability and illumination of the FP may have caused short term deviations. Panel (b) highlights a $\sim2$ month period without significant deviations.
\label{fig:velocities}}
\end{figure*}

We traced the drifts of each of the mode centroids over a period of $\sim6$ months. For comparison with other systems and as a simple examination of the FP stability, we measure the median drift behavior of all the measured resonant modes, shown in Figure \ref{fig:velocities} as a velocity shift. By taking this median, we collapse the wavelength-dependent behavior, which is explored in more detail below. The long-term behavior of the median shift is smooth, although a number of short-term deviations are apparent. Many of these are related to periods of instability in the environment or illumination of the FP; we defer the detailed study of these short term variations for a future work.

\begin{figure}
    \includegraphics[width=0.5\textwidth]{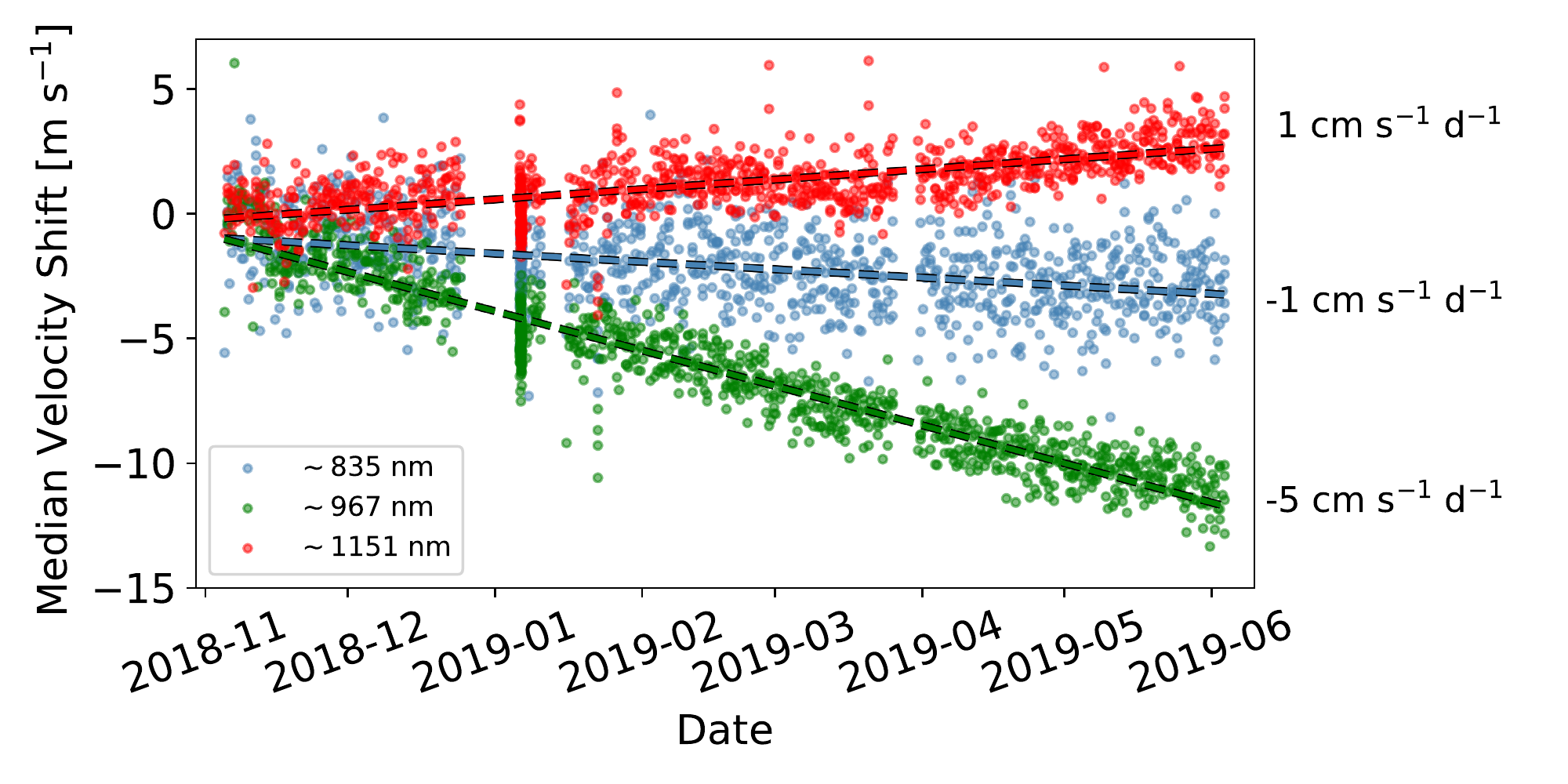}
    \caption{Long-term drift behavior of three sets of 30 FP modes, grouped around 835 nm, 967 nm, and 1151 nm, showing examples of spectrally-resolved drifts, in contrast with the median drift shown in Figure \ref{fig:velocities}. Each group is near the center of an HPF spectral order. Measured centroid shifts for individual etalon modes, as measured via the Gaussian fitting technique shown in Figure~\ref{fig:fits}, are shown as Doppler shifts. The range of behavior spans Doppler shifts of varying signs and magnitudes. The long-term behavior can be approximated as linear in each case. The variation in scatter for each group results from the S/N with which these centroids were measured (see Figure \ref{fig:snrs}).}
    \label{fig:drift_examples}
\end{figure}

However, considering only the median velocity shift belies the wide spectral variation in long-term drift behavior we have measured. A few examples showing the range of these drifts are presented in Figure \ref{fig:drift_examples}, represented again as velocity shifts. The spectral variation in both sign and magnitude of the measured velocity shifts is surprising, and suggests an origin beyond a simple long-term change in the FP mirror separation $L$ as discussed in Section \ref{sec:fpdesc}.

\begin{figure*}
    \plotone{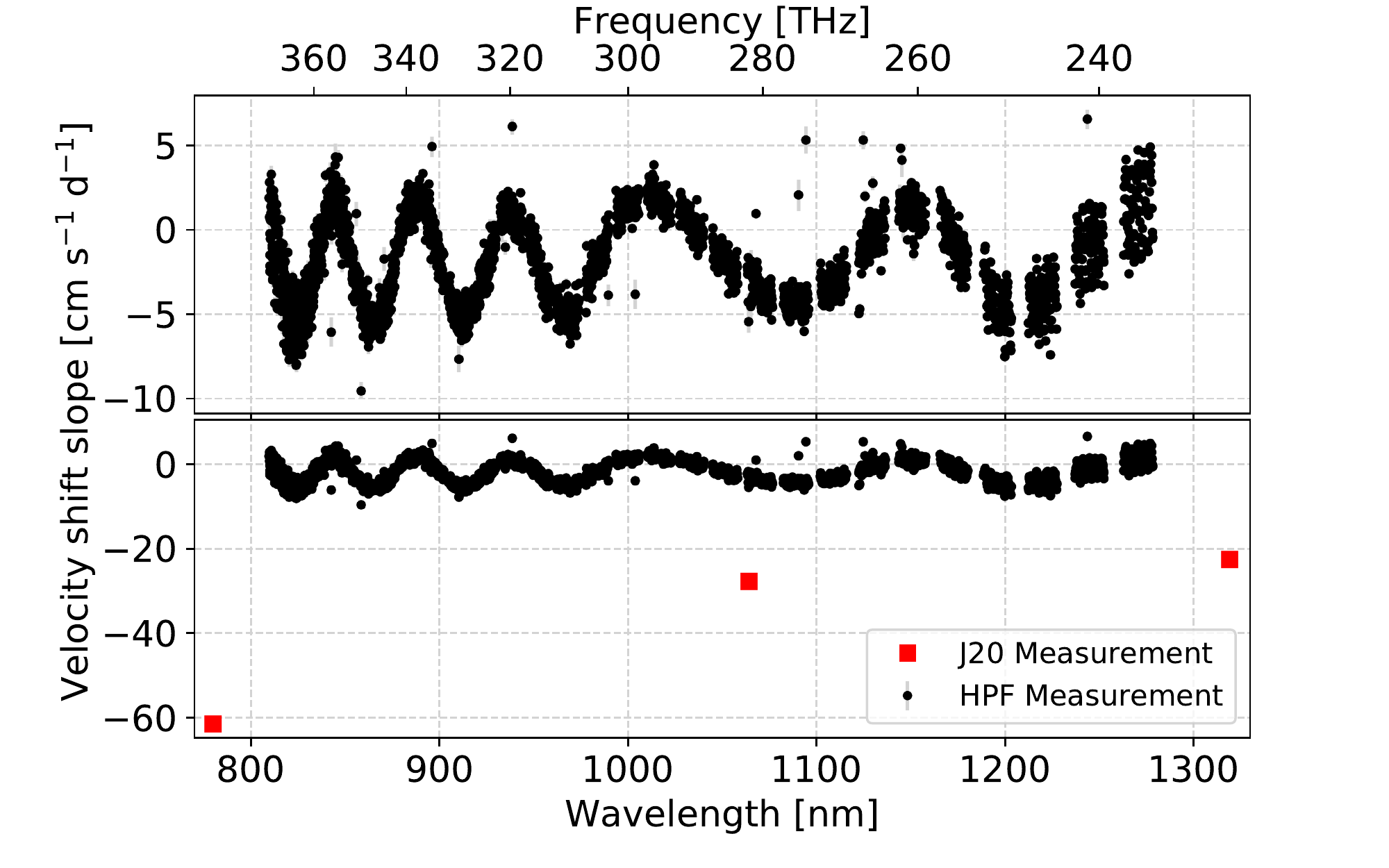}
    \caption{(Top:) Fitted linear slopes of the FP mode drifts over $\sim1000$ epochs from November 2018 through May 2019. The long term linear drifts show a clear wavelength-dependence, and vary in both sign and magnitude. (Bottom:) The same measurements, but plotted along with the results from J20, showing the significantly smaller magnitude of the measured drifts of this system in its present configuration. }
    \label{fig:slopes}
\end{figure*}

In order to probe the nature of these wavelength-dependent (chromatic) drifts, we note that the long-term drift is generally linear, and fit for the slope (m s$^{-1}$ d$^{-1}$) at each wavelength, shown in Figure \ref{fig:slopes}. A clear wavelength-dependent pattern is apparent, with the drift rates varying in both sign and magnitude across the spectrum. Figure \ref{fig:slopes} also shows the measured drift rates of this FP from J20, and makes clear that current measurements of the drift rates for this FP are substantially smaller; this difference is discussed in Section \ref{sec:jenningsdiff}. 

\section{Discussion} \label{sec:discussion}

\subsection{Bulk Shift}
\label{sec:bulkshift}
\begin{figure}
    \includegraphics[width=0.51\textwidth]{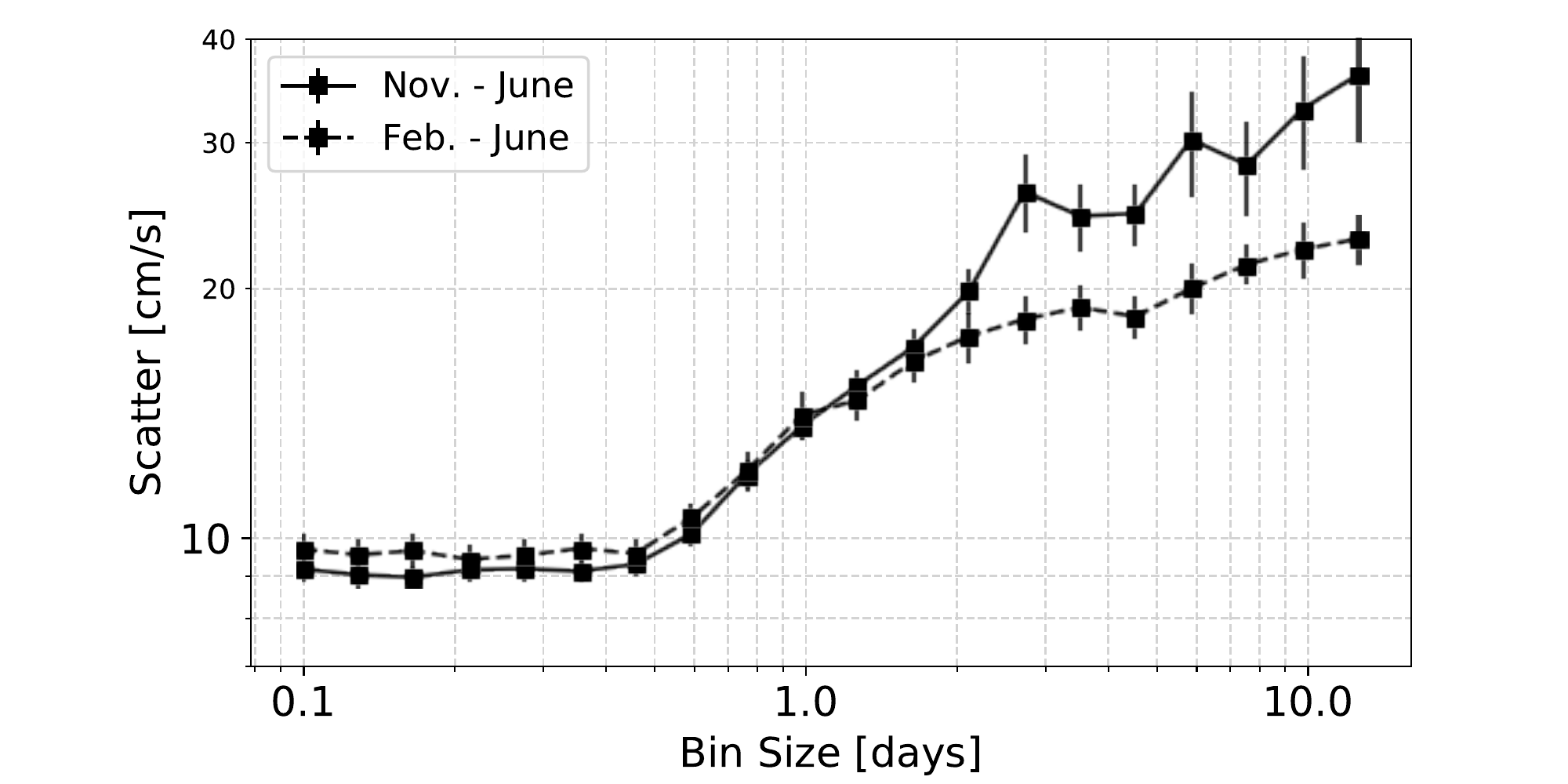}
    \caption{Measured scatter in the median velocity shift for the entire FP spectrum for a range of bin sizes over the stable period identified in Figure \ref{fig:velocities}, showing the stability of the spectrum as a calibrator in the case where spectrograph drift is dominated by a velocity-like term. The dashed line shows only the second half of our full date range, which emphasizes a period where the FP drift behavior was somewhat smoother.}
    \label{fig:binned_stability}
\end{figure}

The median drift of the FP during the study period amounts to around 4 {\ms} over 7 months ($\approx 2$~cm~s$^{-1}$~d$^{-1}$), which compares favorably with similar systems \citep[][]{2020arXiv200601684B,2017JATIS...3b5003S}. To estimate the performance of this system as a calibrator for pure velocity shifts, we explore the scatter of the median drift of the FP at different timescales using the 2-month period of stable operation indicated in Figure \ref{fig:velocities}. We bin the median velocity shifts at varying times and calculate the resulting standard deviations (sigma-clipping at 5-sigma). The results, shown in Figure \ref{fig:binned_stability}, indicate that this scatter within a night is $\lesssim 10$~{\cms}, and $\lesssim 30$~{\cms} over several days. This can be understood as the expected quality of the calibration provided by the FP in the case where spectrograph drift is dominated by a velocity-like term. These results are promising for the use of FP systems as calibrators in this context, but we caution that there are still unexplained short and long-term variations in the FP drift behavior that are under study.

\subsection{Difference with Jennings et al.~2020 measurement}
\label{sec:jenningsdiff}
The bottom panel of Figure \ref{fig:slopes} shows that the present drift rates differ significantly from those measured in J20. These latter were based on repeated temporal scans of three FP modes (at 780, 1064, and 1319~nm) with tunable continuous-wave lasers. The FP transmission profiles were recorded, and the laser wavelength at any given time was recorded by measuring the heterodyne beat frequency generated when the laser was mixed with a stabilized, self-referenced 250 MHz frequency comb system.  

The origin of the difference between the present measurements and those from J20 is presently unknown, but there are several important differences between the two measurements. One is that the J20 measurements were performed during early 2018, approximately one year before those presented here, and closer in time to when the system was assembled and placed in vacuum for the first time. Stabilized FP etalons are known to undergo some relaxation related to the spacer materials and stresses in the optical contacts between the mirrors and the spacer \citep{2009ApPhB..95...43D,1977Metro..13....9B}. The result is typically that the overall length of the FP etalon shrinks over time, with an asymptotically decreasing shrink rate. Further, between the J20 measurements and those presented here, the system was shipped from the NIST lab in Colorado to the HET in Texas. It is possible that during this shipment, the system underwent a kind of ``annealing" in which mechanical stresses were rapidly dissipated. A combination of the FP relaxation and potential mechanical ``annealing" during shipping could explain the differences in drift rates measured.

It is also important to compare the two measurement techniques, both of which achieve excellent absolute accuracy by referencing to stabilized LFCs. The J20 measurements achieve high sampling and precision on the centroids of three individual FP resonant modes, and involve illuminating the FP only selectively with the scanning lasers. The present measurements are much less accurate for any individual mode, but leverage an ensemble of $\sim3500$ modes across the HPF bandpass, and involve illuminating the FP across its full bandpass (including the supercontinuum pump laser, which contains substantial optical power). These two techniques are therefore subject to different systematic errors. For example, environmentally-sensitive low-amplitude parasitic etalons might hamper the centroid measurement of an individual FP mode in a way that averages out over larger groups of nearby modes. On the other hand, the present measurements may be subject to effects from the varying optical power in the supercontinuum pump laser which are not present in the J20 measurements.

\subsection{The origins of the wavelength-dependent drift} \label{sec:wavelengthdrift}
The most important result of this study is the specific wavelength-dependent FP drift behavior observed in Figure~\ref{fig:slopes}. To our knowledge, this is the most detailed long-term chromatic study of a broadband astronomical FP yet published. Importantly, the observed behavior deviates from the commonly-held assumption that the long-term drift behavior of such an FP is dominated by a slow relaxation of the FP spacer material and interfaces, resulting in a gradual shrinkage of the mirror separation (and shifting of the modes towards higher frequencies/smaller wavelengths). Indeed, the \textit{average} drift measured here (Figure \ref{fig:velocities}) is small in magnitude (a few cm~s$^{-1}$~d$^{-1}$) and in the appropriate direction (blueshift), and is similar to the drift measured in \citet{2017JATIS...3b5003S} that is attributed to the slow material relaxation. However, the present measurements confirm the result of J20 that the drift of the FP modes is not related purely by their mode indices (as described in Section \ref{sec:fpdesc}), and therefore is not driven by a change in the cavity mirror separation $L$ alone.

There are a number of possible physical origins for this effect. Noting the oscillatory nature of the variations in Figure \ref{fig:slopes}, one possibility is a low-amplitude ``parasitic" etalon introduced by reflections within the optical path of the FP system. Such an etalon might be introduced by, for example, reflections within the mirror substrates, or off fiber terminations, or within the Bragg grating notch filter. Thus exposed to environmental or other variations, the effect on the measured spectrum might amount to a gradually-changing ``peak-pulling" effect that manifests as the variations we measure. 

A single such parasitic etalon (or its peak-pulling effect) would (to first order) be described by Equation \ref{eq:FSRsimple}. To the extent that the material refractive index is constant, the {\fsr} should be roughly constant, and so any peak-pulling effect should also be roughly periodic in frequency. For reference, a parasitic etalon with {\fsr}$\approx20$~THz would correspond to $L\approx10$~$\mathrm{\mu}$m, and a parasitic etalon ``beating" with the FP resonant modes at a frequency of $\sim20$~THz would correspond to an $L$ very nearly equal to the few mm spacing of the FP itself. Inspection of the frequency axis in Figure \ref{fig:slopes} reveals that the period of the effect varies by a factor of three. For a single parasitic etalon this implies a change in the {\fsr} by a similar factor, which we consider unlikely. We therefore consider the impact of a single parasitic etalon (resulting from, for example, the Bragg grating) to be an unlikely explanation. A more complicated combination of such effects cannot be ruled out, however.

Another possibility relates to how the in-coupled light is incident on the FP. The FP is illuminated by broadband supercontinuum light, which is injected through a single-mode optical fiber and collimated by an off-axis parabolic mirror. The numerical aperture and mode pattern exiting the input fiber is wavelength-dependent, and this could in principle translate to different distributions of incident angles onto the FP. Coupled with long-term mechanical settling of the optical mounts, this might yield changes that manifest as the measured chromatic drifts. A non-normal angle of incidence $\theta$ onto an etalon can be parameterized by simply substituting $L\,\cos \theta$ for $L$ in Equation \ref{eq:fsr}. A change in $\theta$ with this simple parameterization would cause a uniform offset of the {\fsr} values, which would manifest as a (spectrally) smooth change in {\fsr} or Doppler shift. The linear changes shown in Figure \ref{fig:slopes} are clearly not well-described by such a change, and producing the observed pattern with an incident angle mechanism would appear to require a complicated illumination pattern that varies non-monotically with wavelength. We consider that change in the angle-of-incidence onto the FP is therefore unlikely to be the main cause of the chromatic drift behavior observed.

Other potential explanations for this drift behavior involve the FP mirror coating itself. We have shown that phase delays resulting from this coating are likely playing a measurable role is determining the FP mode spectrum (Section \ref{sec:static}), so it is possible that changes in this mirror coating are driving changes in the FP mode spectrum as well. The mechanism of this hypothesized effect could involve changes in either the thicknesses or the refractive indices of the layers in the dielectric stack. An exploration of the details of this type of effect will be important for refining the design of future FP systems used for long-term high-precision Doppler spectroscopy, but is beyond the scope of this work.

Although we do not aim for a comprehensive list of potential  ``ultimate'' causes for the changes described above, we note that the incident optical power to the FP from the supercontinuum source is variable, being turned off and on regularly to extend the usable lifetime of the FP system. When active, the supercontinuum source illuminates the FP with $\sim1$~mW of optical power, and this could, in principle, impose thermal transients on downstream optical components (e.g., the FP mirrors). 

Tests of such optical ``heating'' with a narrower-band illumination at similar power \citep{Jennings2020} suggest that $\lesssim1$~mW illumination does not impact the mode centroid positions at the few 100~kHz level. However, as noted in Section \ref{sec:jenningsdiff}, it is not clear how equivalent the present FP configuration at the HET is to the testing configuration of \citet{Jennings2020}. These configurations differ in their illumination source profile, the timing of the illumination, and potentially in the optical alignment of elements in the FP system. It is also non-trivial to accurately determine the spectral profile of the resonant optical power in the FP in the presence of the evolving spectral profiles of the illumination source and the FP itself. Thus, while we cannot yet trace the impact of optical heating in our measurements, this effect remains an important area of investigation for this system.

Finally, it is important to note the potential role of subtle changes in the HPF instrumental profile (IP) in the precise, long-term monitoring of the FP drift. Even in a highly stabilized spectrograph like HPF, the IP is expected to evolve over the long term. This evolution could, in principle, drive the appearance of wavelength-dependent centroid drifts in the FP monitoring because this analysis assumes a simple and unchanging line profile.

Indeed, information from the HPF LFC and FP have been central to measuring the long-term drift of the HPF IP, and the dominant terms of the drift correction address a slow evolution in the camera magnification and a relaxation of the HPF inter-pixel capacitance (Terrien et al.~in prep), which both appear as IP changes in the non-drift corrected data. Importantly, these effects in the non-drift corrected data are highly correlated with the position of a particular wavelength in the HPF echelle format, due to the dependence of optical aberration or detector behavior on focal plane position. Because no such correlation remains in the wavelength-dependent drifts we present here (e.g.~Fig.~\ref{fig:slopes}), we consider it unlikely that the evolution of the IP is the source of these drifts. We cannot rule out, however, \textit{all} potential IP changes, and efforts are underway to leverage the unique flexibility of the LFC to measure these changes directly.

\subsection{Implications for spectrograph calibration}
The chromatic structure in the behavior of this FP system has significant implications for the calibration of Doppler spectrographs that aim at $\sim10$~{\cms} precision. With a FP calibrator, such systems may indeed by limited by the interplay of these chromatic variations, and other instrumental systematics related to the detector or spectrograph drift. Our results show that the extraction of the highest precision Doppler shifts will require extensive characterization of the spectral and temporal characteristics of the spectrograph and the FP calibration system. The present FP calibration system clearly does not conform to the expectation that the time evolution of the FP mode spectrum can be predicted simply by projecting the evolution of a single mode (or a small number of modes) under the assumption of the dominance of the change in mirror separation $L$, so we caution against relying on this assumption in other systems. 

At the level of $\sim1$~{\ms}, the results in Section \ref{sec:bulkshift} show that this FP system performs well as a standalone calibrator for time periods up to at least $\sim10$~days. This system is in use as a ``holdover" calibrator at HPF for times when the LFC is unavailable, and is used to derive the drift model parameters as explained in Section \ref{sec:FPwavecal}. We note that the drift behavior of this FP and of HPF are sufficiently independent that the process of refining the FP mode measurements resulted in the discovery and subsequent implementation of the linear and readout channel terms in our HPF drift model.

Our results also emphasize the need for similar in situ characterization of other systems, in view of the significant differences between our present findings and those presented in J20. We also note that there are a wide variety of astronomical FP calibrator designs in use already, many of which do not (yet) have the benefit of an accompanying LFC to evaluate their behavior in situ. A non-exhaustive list of the differences in design choices among these systems includes different illumination sources (laser-driven light sources, supercontinuum sources), different coupling (single-mode, multi-mode), different resonator spacer and mirror substrate materials (ULE, Zerodur), and different mirror coating properties (soft or hard coatings, low or high finesse). This variety in the face of a similar goal---wavelength calibration of high-resolution Doppler spectrographs---suggests that these designs can be optimized and improved. The results presented here are pertinent to our specific design; further study will be necessary to establish the extent to which other designs are subject to this type of behavior, but this work serves as a cautionary note about assuming a simple relationship relating the drift of different FP modes.

\subsubsection{Calibration of HPF using the FP}
\label{sec:FPwavecal}
The HPF FP has been used as a ``holdover" calibration source for brief (one to a few days) periods when the LFC is unavailable. Here, we briefly discuss our current method for using the FP in this context. 

As described in Section \ref{sec:wavecal}, our HPF drift model comprises a second-order polynomial (per order), along with discrete offsets for each read-out channel. In the short term (days to weeks), the drift is dominated by the $0^{\mathrm{th}}$-order term (i.e., pixel offset) of the polynomial, and the higher-order terms vary more slowly on the timescale of weeks to months.

Due to the wavelength-dependent drift of the FP we uncovered in this study, as well as the small time baselines when the FP is needed, we presently only rely on the FP to estimate the $0^{\mathrm{th}}$-order term of the polynomial. To estimate the value of this term at a given epoch, we follow a similar procedure to that described in Section \ref{sec:wavecal}. We first create a high S/N FP template spectrum by combining multiple epochs of wavelength-calibrated FP spectra. This template spectrum is then fit via least-squares optimization to a target epoch, with a single pixel offset as the only free parameter. The resulting pixel shift term constrains the 0$^{\mathrm{th}}$-order term of the drift model for a single point in time. 

\begin{figure}
    \includegraphics[width=0.5\textwidth]{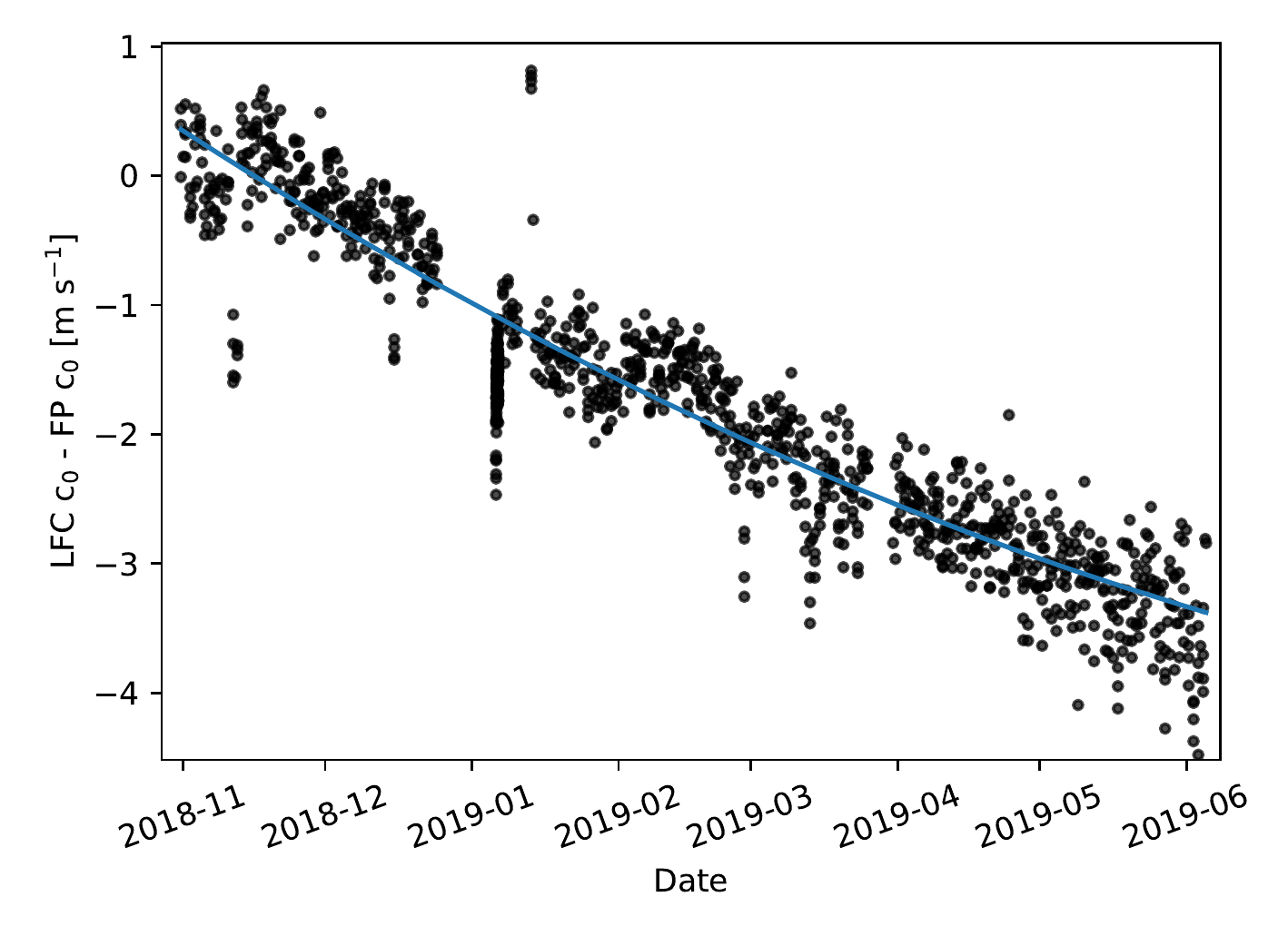}
    \caption{The difference between the 0$^{\mathrm{th}}$ order term in the LFC drift model, and the corresponding term estimated from the FP spectrum. As in Figure \ref{fig:velocities}, there are a small number of times when environmental or illumination changes induced short-term deviations. A second-order polynomial approximation which is used to convert from pixel drift from the FP measurement to the  0$^{\mathrm{th}}$-order term in the drift model is shown by the blue curve. For clarity, the $y$-axis is shown in units of  velocity ({\ms}) instead of the scaled pixel units.}
    \label{fig:etalonc0versusLFCc0}
\end{figure}

This constraint is subject to a slow drift, as indicated by the FP drift measurements in Figure \ref{fig:velocities}. Figure \ref{fig:etalonc0versusLFCc0} shows the difference between the 0$^{\mathrm{th}}$-order term in the drift model (LFC c$_0$) and that derived directly from the FP (FP c$_0$), as measured in our wavelength calibration. A smooth (second-order) polynomial fit (also shown in Figure \ref{fig:etalonc0versusLFCc0}) is sufficient to correct the FP-based offset measurement for this long-term drift. 

\section{Conclusion} \label{sec:conclusion}
Broadband {\fap} etalons offer an attractive alternative to classical RV calibration sources, with a rich spectrum of sharp features at regular frequency spacing. These devices have many distinct advantages over atomic emission lamps and molecular absorption cells, and are intrinsically significantly simpler than the ``gold-standard" of a laser frequency comb. However, their use for calibration at the $\sim10$~{\cms} level, and as long-term calibration sources in their own right, remains largely untested.

We performed an extensive spectral study of the stability of a {\fap} etalon calibration system for the HPF spectrograph. We leveraged the long-term accuracy and precision provided by the dedicated HPF LFC calibration system. Using the LFC-based wavelength calibrations, we located and tracked the centroids of each of $\sim3500$ {\fap} resonant modes (Figure \ref{fig:fits}) over a period of six months, confirming that we were photon-limited at any one epoch (Figure \ref{fig:snrs}). Our main results for this FP system are:
\begin{itemize}
    \item Chromatic differences in the single-epoch mode spectrum of the FP are traceable to the resonator mirror coating (Figure \ref{fig:fsrs}).
    \item The median drift of the FP supports a calibration precision of $\lesssim 30$~{\cms} over $\lesssim10$~days and $\lesssim 10$~\cms over a night (Figures \ref{fig:velocities}, \ref{fig:binned_stability}).
    \item The drift rate of modes across the FP spectrum is a complicated function of wavelength, varying in both sign and magnitude (Figures \ref{fig:drift_examples}, \ref{fig:slopes}). These drift rates do not conform to the expectation of a simple model (a gradually shrinking resonator mirror separation) that has been suggested to explain long-term drifts in other systems.
\end{itemize}
We discussed potential origins for this chromatic behavior, including parasitic etalons, illumination angle changes, and changes in the FP mirror coating. We also discussed potential origins for the significant differences from measurements of this system presented in J20.

These measurements represent an important step forward in characterizing {\fap} etalons for precise RV spectrograph calibration. Our results highlight the need for extensive characterization of FP systems over long timescales, and suggest that a multi-wavelength study of the FP stability in situ may be required for enabling the highest levels of RV precision.

\acknowledgments
The authors wish to thank the anonymous referee for insightful comments which substantially improved this work. The authors also wish to thank Light Machinery for graciously sharing the resonator mirror phase curve data. The authors also thank Mark Notcutt and Mike Grisham at Stable Laser Systems for sharing their knowledge about FP systems and their insight into the design of this system. RCT thanks Bruce Duffy for his advice and support with managing computational resources at Carleton. 

This work was partially supported by funding from the Center for Exoplanets
and Habitable Worlds. The Center for Exoplanets and Habitable
Worlds is supported by the Pennsylvania State University, the
Eberly College of Science, and the Pennsylvania Space Grant
Consortium. These results are based on observations obtained with
the Habitable-zone Planet Finder Spectrograph on the HET. We
acknowledge support from NIST and the NIST-on-a-Chip Program, NSF grants AST 1006676, AST
1126413, AST 1310875, AST 1310885, ATI 1006676, ATI 2009889, ATI 2009982, and the NASA
Astrobiology Institute (NNA09DA76A) in our pursuit of precision
radial velocities in the NIR. We acknowledge support from the
Heising-Simons Foundation via grant 2017-0494. The HET is a
joint project of the University of Texas at Austin, the Pennsylvania
State University, Ludwig-Maximilians-Universität München, and
Georg-August Universität Gottingen. The HET is named in honor
of its principal benefactors, William P. Hobby and Robert E.
Eberly. 

The HET collaboration acknowledges the support and resources from the Texas Advanced
Computing Center. We thank the Resident Astronomers and
Telescope Operators at the HET for the skilful execution of our
observations of our observations with HPF.

Computations for this research were performed on the Pennsylvania State University’s Institute for Computational and Data Sciences’ Roar supercomputer,
including the CyberLAMP cluster supported by NSF grant MRI1626251.

\vspace{5mm}
\facilities{HET(HPF)}

\software{\texttt{astropy} \citep{2013A&A...558A..33A,2018AJ....156..123A}, 
\texttt{numpy} \citep{5725236},
\texttt{scipy} \citep{2020NatMe..17..261V},
\texttt{HxRGproc} \citep{2018SPIE10709E..2UN},
\texttt{matplotlib} \citep{4160265},
\texttt{GNU Parallel} \citep{tange2011gnu},
          }

\newpage
\bibliography{ads_20200703}{}
\bibliographystyle{aasjournal}

\end{document}